\documentclass[twocolumn]{aa}
	\usepackage[bookmarks=false]{hyperref}
	\usepackage{amsmath,amssymb,mathrsfs}
	\usepackage{graphicx,natbib,verbatim}
	\usepackage[varg]{txfonts}

	\bibpunct{(}{)}{;}{a}{}{,}

\begin{document}

	\titlerunning{Non-Gaussian solar flare spectral line profiles}
	\authorrunning{Jeffrey, Fletcher \& Labrosse}

	\title{First evidence of non-Gaussian solar flare EUV spectral line profiles and accelerated non-thermal ion motion}

	\author{Natasha L. S. Jeffrey, Lyndsay Fletcher \and Nicolas Labrosse}

	\offprints{{N. L. S. Jeffrey \email{natasha.jeffrey@glasgow.ac.uk}}}

	\institute{School of Physics \& Astronomy, University of Glasgow, Glasgow G12 8QQ, UK}

	\date{Received ; Accepted}

	\abstract{The properties of solar flare plasma can be determined from the observation of optically thin lines. The emitting ion distribution determines the shape of the spectral line profile, with an isothermal Maxwellian ion distribution producing a Gaussian profile. Non-Gaussian line profiles may indicate more complex ion distributions.}
	{We investigate the possibility of determining flare-accelerated non-thermal ion and/or plasma velocity distributions.}
	{We study EUV spectral lines produced during a flare SOL2013-05-15T01:45 using the Hinode EUV Imaging Spectrometer (EIS). The flare is located close to the eastern solar limb with an extended loop structure, allowing the different flare features: ribbons, hard X-ray (HXR) footpoints and the loop-top source to be clearly observed in UV, EUV and X-rays. EUV line spectroscopy is performed in seven different regions covering the flare. We study the line profiles of the isolated and unblended Fe XVI lines ($\lambda262.9760~$\AA\,) mainly formed at temperatures of $\sim$2 to 4 MK. Suitable Fe XVI line profiles at one time close to the peak soft X-ray emission and free of directed mass motions are examined using: 1. a higher moments analysis, 2. Gaussian fitting, and 3. by fitting a kappa distribution line profile convolved with a Gaussian to account for the EIS instrumental profile.}
	{Fe XVI line profiles in the flaring loop-top, HXR footpoint and ribbon regions can be confidently fitted with a kappa line profile with an extra variable $\kappa$, giving low, non-thermal $\kappa$ values between 2 and 3.3. An independent higher moments analysis also finds that many of the spectral line kurtosis values are higher than the Gaussian value of 3, even with the presence of a broad Gaussian instrumental profile.}
	{A flare-accelerated non-thermal ion population could account for both the observed non-Gaussian line profiles, and for the Fe XVI `excess' broadening found from Gaussian fitting, if the emitting ions are interacting with a thermalised $\sim$4 MK electron population, and the instrumental profile is well-approximated by a Gaussian profile.}


	\maketitle

	\section{Introduction}\label{intro}
	When a solar flare occurs, a portion of the released energy goes into accelerating particles. The accelerated particles are transported within, interact with and create hot Megakelvin plasma, the properties of which are mainly investigated by the radiative signatures of soft X-rays (SXR) and (extreme) ultraviolet (EUV/UV) continuum and line emissions \citep[e.g.][]{2011SSRv..159...19F}. Spatially resolved EUV line spectroscopy of optically thin lines is currently performed using the Extreme Ultraviolet Spectrometer (EIS) \citep{2007SoPh..243...19C} onboard {\em Hinode\,} and for certain optically thin lines in the UV range, with the Interface Region Imaging Spectrograph (IRIS) \citep{2014SoPh..289.2733D}. Flares are non-equilibrium processes, but most flare spectral line studies extract information under the conditions of ionization equilibrium and local thermal equilibrium. The spectral line profiles are usually well-fitted (or at least well-approximated) with a Gaussian and the properties of the flaring plasma are extracted using the first three normalised moments: the ion abundance and electron density from the integrated intensities (zeroth moment), directed plasma motions from shifts about the centroid positions (first moment) and temperature and/or random plasma motions from the line broadening (second moment). In a hot, flaring solar atmosphere, the spectral lines are expected to be dominated by Doppler broadening; the spectral line profile is broadened by many small shifts in wavelength, caused by the random, isotropic motions of the emitting ion distribution along the line of sight. Doppler broadening produced by ions with a Maxwellian velocity distribution creates a Gaussian line profile, and the line width is proportional to the square root of the ion temperature, usually taken to be the peak contribution temperature of the line. Other broadening mechanisms can produce non-Gaussian line profiles, namely increased collisions in high density regions (collisional broadening) leading to broad-winged Lorentzian line profiles. However, in the majority of solar flare cases, collisional broadening should be negligible ($\Delta\lambda\sim10^{-15}~$\AA) compared with Doppler broadening, even for electron densities of $10^{11}$ cm$^{-3}$ \cite[for more information, see][]{2011ApJ...740...70M}. 

	The majority of observed solar flare spectral lines show an excess broadening, where the measured line widths are larger than those expected from ion thermal motions alone. The cause of excess broadening has been debated for years, and the most common explanations are given by turbulence (random plasma motions) along the line of sight \citep[e.g.][]{1995ApJ...438..480A,1993SoPh..144..217D,1979ApJ...233L.157D,1980ApJ...239..725D,1990A&A...236L...9A,1986ApJ...301..975A}. The excess broadening is usually associated with a single non-thermal velocity estimated from the excess (and also assumed Gaussian) width. A velocity distribution of random fluctuations due to plasma waves, for example, could produce non-Gaussian line shapes as well as excess broadening. Microscopic deviations from an isothermal Maxwellian ion distribution could also produce non-Gaussian line shapes, and hence the excess broadening could be produced by an isotropic but accelerated non-thermal ion population, as suggested by \citet{1992ApJ...398..319S}, particularly during a flare. \citet{2008ApJ...679L.155I} found evidence of non-Gaussian line profiles during an X class flare, finding that broad non-Gaussian profiles were associated with red-shifts in the flare arcade. A study by \citet[][published but not yet peer-reviewed]{2013arXiv1305.2939L} investigated the shapes of Fe XV line profiles in the non-flaring solar corona. Their analysis suggests that the lines at non-flaring times were fitted better by a kappa distribution line profile controlled by an extra parameter $\kappa$ \citep[cf][]{2009JGRA..11411105L}. Non-thermal kappa-distributed heavy ion populations are ubiquitous in the collisionless solar wind and are routinely detected \citep[e.g.][]{1998SSRv...86..127G}. During a flare, it is likely that the ions are also excited by flare-accelerated, non-thermal electrons, as well as a Maxwellian electron distribution. In the last few years, many studies \cite[e.g.][]{2014A&A...570A.124D,2015ApJS..217...14D}, have recalculated the continuum and line emissions produced by an ionising kappa distribution of electrons, that can account for the presence of a power-law tail of high energy accelerated electrons. Since the electrons are responsible for the formation of a line, the velocity distribution(s) of the ions, whether thermal or accelerated is unimportant for many studies. However, non-Gaussian line profiles could provide a valuable plasma diagnostic for the determination of flare-accelerated non-thermal ion motions, and the underlying flare processes. Finally, a non-Gaussian line profile might also provide information about the multi-thermal nature of the flaring plasma. 

	In this paper, we study the line profiles of Fe XVI in seven different regions of a solar flare, SOL2013-05-15T01:45, at one time close to the flare SXR peak. Section \ref{flare} discusses the observation of SOL2013-05-15T01:45 using EIS. Section \ref{profiles} presents the evidence for non-Gaussian line profiles in different regions of the flare using two different techniques of (1) a higher moments analysis, and (2) line fitting. Section \ref{errors} discusses instrumental issues and the range of detectability and uncertainty associated with measuring non-Gaussian line profiles with EIS. Finally, in Section \ref{discuss}, we discuss the possible causes of non-Gaussian line profiles and excess broadening, in particular the possibility of flare-accelerated non-thermal ion populations.

	\begin{figure}
	\centering
	\includegraphics[width=7.6 cm]{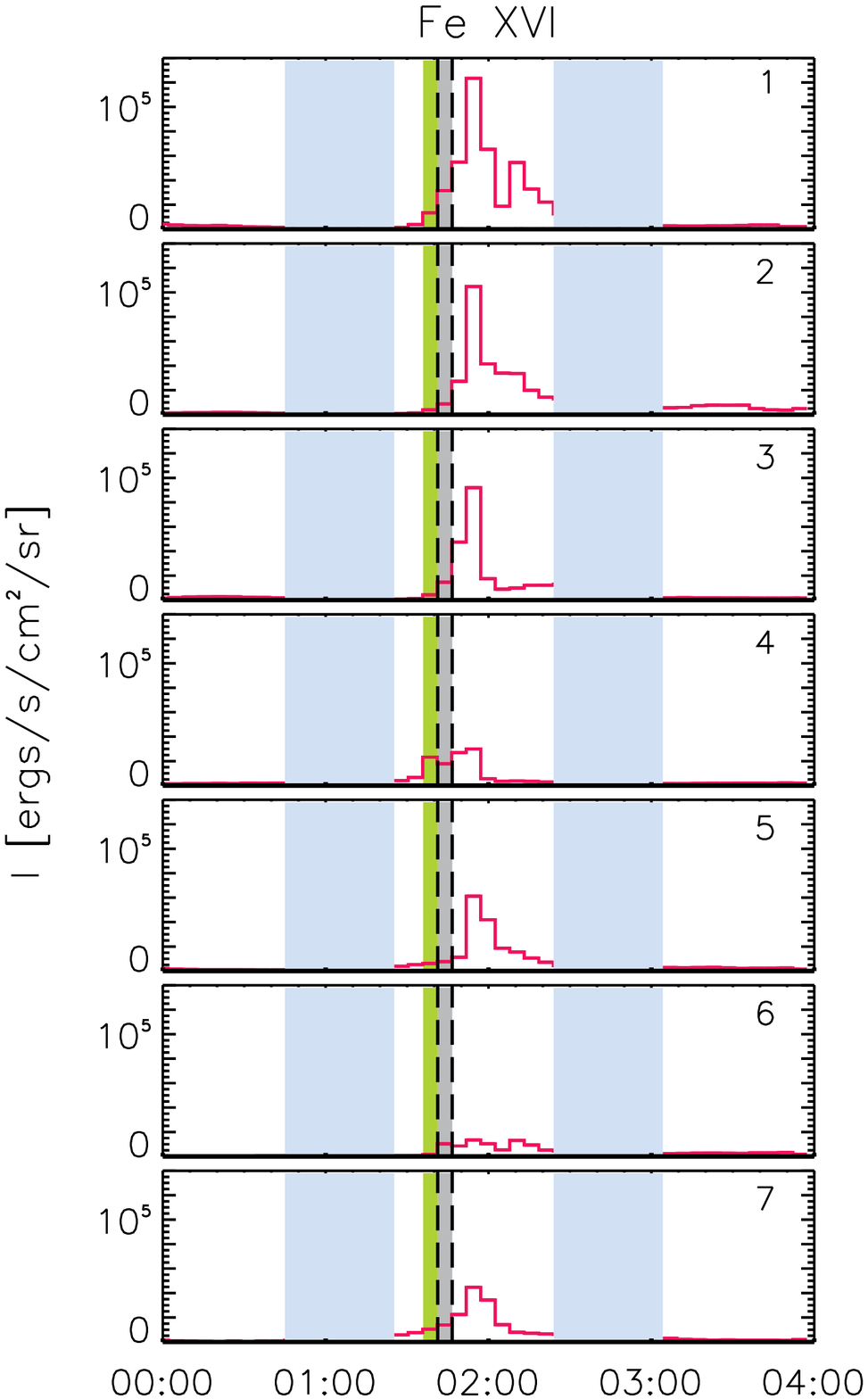}\vspace{0.1cm}
	\includegraphics[width=7.6 cm]{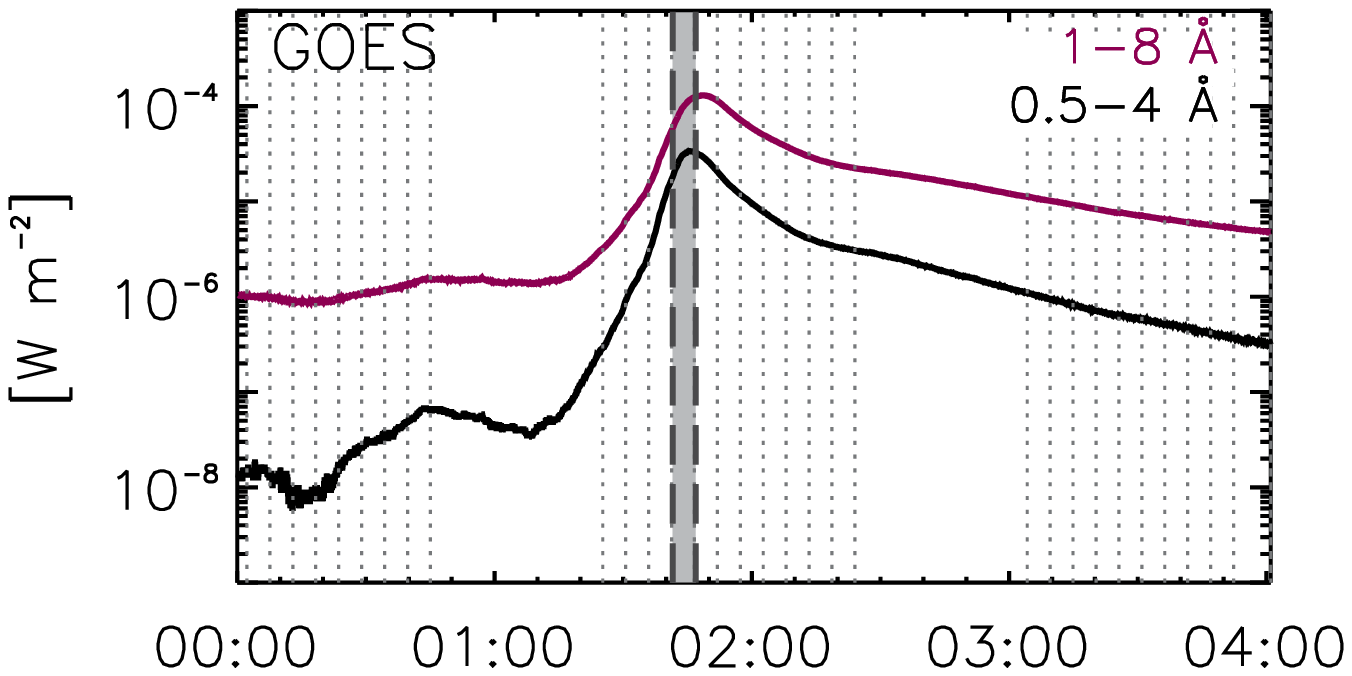}\vspace{0.1cm}
	\includegraphics[width=7.6 cm]{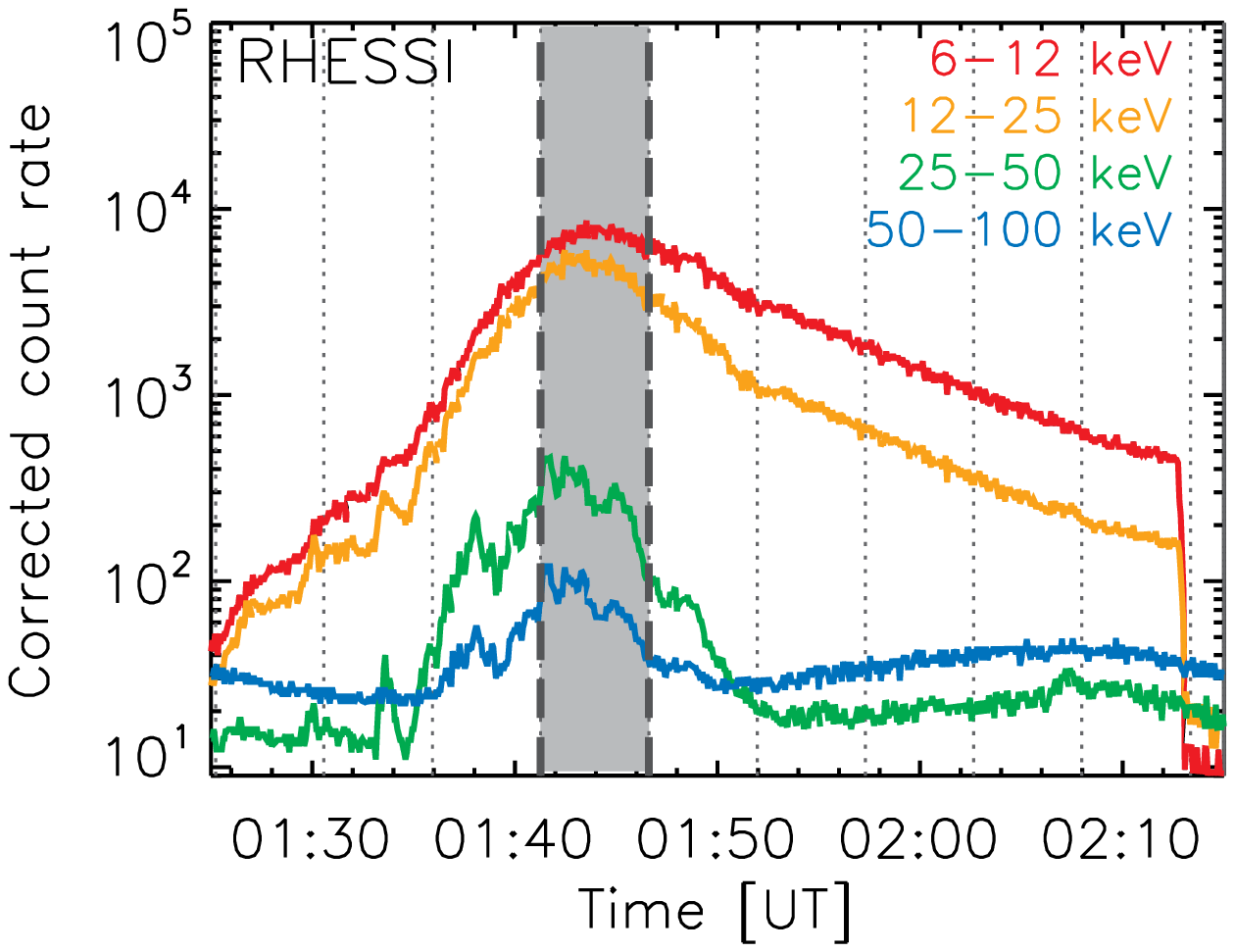}
	\vspace{-0.1cm}
	\caption{{\it GOES\,} (middle) and {\it RHESSI\,} (bottom) lightcurves for SOL2013-05-15T01:45. The {\it GOES\,} lightcurve is shown for a longer period before, during and after the flare. The {\it RHESSI\,} lightcurve is shown for the flare times from $\sim$01:25 to 02:15 UT. The grey dotted lines indicate the start times of an EIS raster observation. The top seven panels show the Fe XVI integrated intensity (1. corona, 2. loop leg, 3. and 4. HXR footpoint, 5. loop leg, 6. HXR footpoint, and 7. ribbon only) of the flare (see Figure \ref{fig3}). Dark grey band - time of study, green band - HXR peak, blue bands - no EIS data.}
	\label{fig1}
	\end{figure}

	\section{The observation of flare SOL2013-05-15T01:45}\label{flare}
	The chosen flare, SOL2013-05-15T01:45, is an X1.2 flare located close to the eastern solar limb. The {\em GOES\,} and {\em RHESSI\,} lightcurves for the flare are shown in Figure \ref{fig1}. The {\em GOES\,} lightcurve shows the SXR flux rising from around 01:00 UT and peaking at $\sim$01:45 UT. The {\em RHESSI\,} lightcurve shows the hard X-ray (HXR) emission above 25 keV starting to rise at 01:34 UT and peaking before the SXR emission at 01:42 UT. In Figure \ref{fig1}, the grey dotted lines indicate the start time of each EIS raster. The EIS observations cover the main flare times from $\sim$01:25 UT to $\sim$02:24 UT, with observations also available before and after the flare times.

\begin{figure}[ht]
	\centering
	\includegraphics[width=0.5\textwidth]{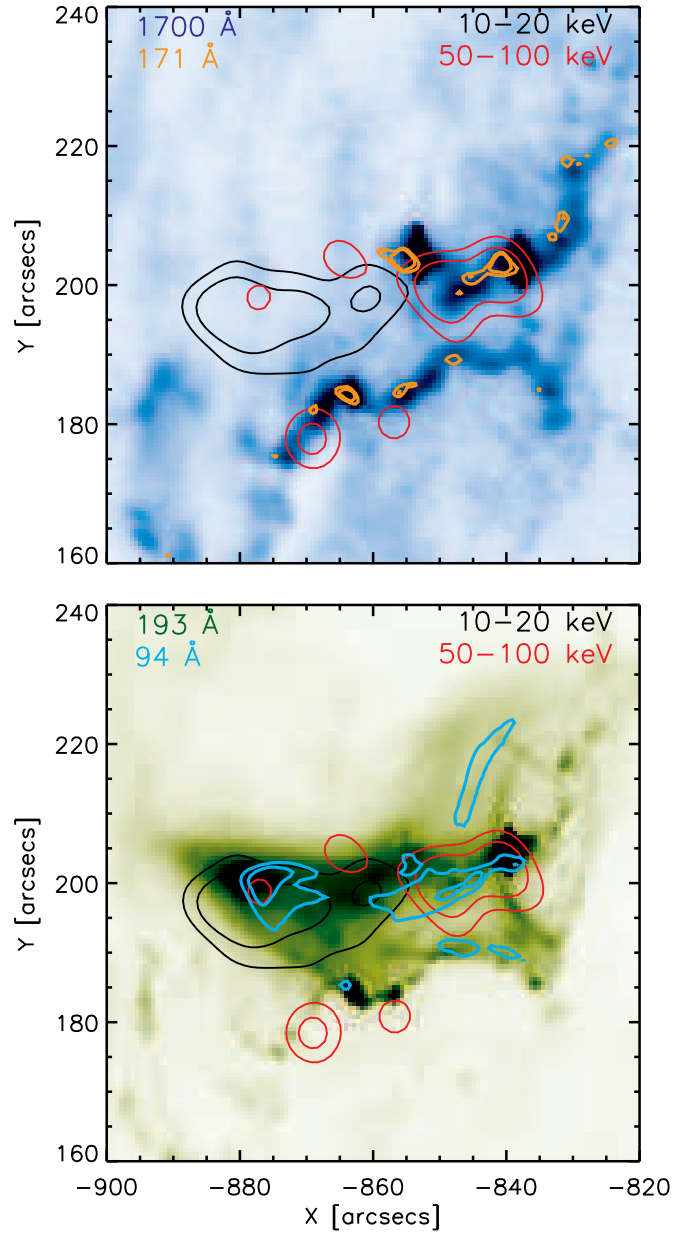}
	\caption{{\em SDO\,} AIA images of SOL2013-05-15T01:45 in 1700 \AA\, (top) and 193 \AA\, (bottom). 1700 \AA\, shows the positions of the ribbons clearly. AIA contours at 171 \AA\, and 94 \AA\, are displayed at 30 \% and 50\% of the maximum. {\em RHESSI\,} X-ray contours are also displayed showing the positions of a loop-top source (black) and hard X-ray footpoints (red) at 30 \% and 50 \% of the maximum. The {\em RHESSI\,} X-ray contours are shown for a time interval of 01:37 to 01:38 UT, while the AIA images are from various times between this interval.}
	\label{fig2}
	\end{figure}

	Images of SOL2013-05-15T01:45 are shown in Figure \ref{fig2} using {\em SDO\,} AIA 1700 \AA\, and 193 \AA\,, at a time interval of 01:37-01:38 UT before the SXR peak. AIA and {\em RHESSI\,} contours at 171 \AA\,, 94 \AA\,, 10-20 keV and 50-100 keV are also displayed on the images. The main features of the flare such as the loop-top source (at 10-20 keV), HXR footpoints (at 50-100 keV), hot loops (at 193 \AA\,) and ultraviolet ribbons (at 1700 \AA\,) can be clearly seen in Figure \ref{fig2}. The eastern limb location of SOL2013-05-15T01:45 and its elongated structure allow the flare features to be clearly observed and hence examined independently of each other without significant overlap between the loop-top source and the ribbons or the HXR footpoints. Since this flare is located at a high heliocentric angle close to the limb ($\sim67^{\circ}$), it is likely that the observer's line of sight is at an angle close to perpendicular to the guiding magnetic field connecting the coronal loop-top to the footpoints. 

	\begin{figure}[ht]
	\centering
	\hspace{-1.1 cm}
	\includegraphics[width=0.54\textwidth]{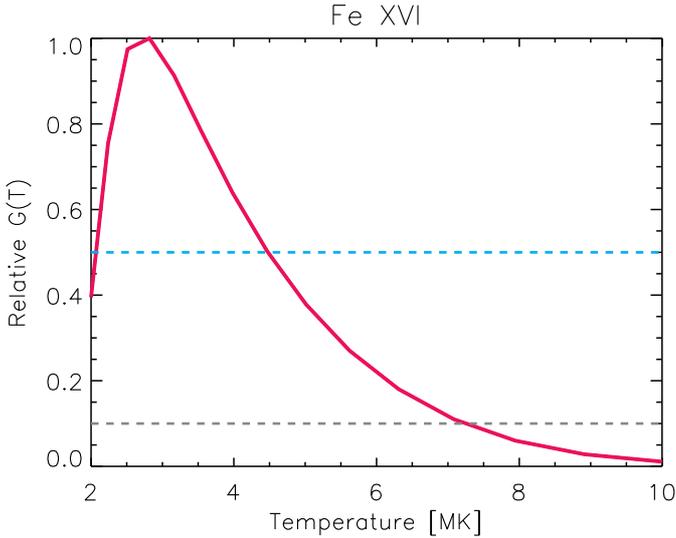}
	\caption{Relative contribution function $G(T)$ (i.e. peak set to 1) for Fe XVI $262.9760~$\AA~ plotted using CHIANTI. Fe XVI is mainly formed over the temperatures of 2 to 4 MK. Over these temperatures, Fe XVI $G(T)$ values are greater than half of the maximum contribution value, showing that Fe XVI is easily formed over this temperature range. The $G(T)$ curve is similar for both coronal and photospheric abundances.}
	\label{fig_goft}
	\end{figure}

	\subsection{Hinode EIS observations of Fe XVI}\label{obs}

	Due to a number of large flares in the preceeding days leading up to SOL2013-05-15T01:45, EIS was already observing the active regions in the area before the start of the flare. During the observation, the two arcsecond slit was used in a fast rastering mode, giving a relatively high temporal resolution of around 9 seconds, but a reduced spatial resolution in the X direction of around $5''.99$ (i.e. the slit position jumps $5''.99$ every $\sim9$ seconds). EIS, when in slit mode, scans a region from solar west to east. In Y, the spatial pixel size is $1''$. The field of view (FOV) covering the flare region is $(30\times5''.99)\sim179'' \times 152''$ arcseconds.

	For a line profile investigation, we need to study strong spectral lines with very little or no blending with other lines. Initially, the iron lines of Fe XII ($195.1190~$\AA), Fe XV ($284.1630~$\AA), Fe XVI ($262.9760~$\AA) and Fe XXIII ($263.7657~$\AA) were studied, but we found that Fe XVI was the best line for a line profile study during the flare. Fe XVI is emitted at a laboratory rest wavelength of $262.9760~$\AA. It is a strong, well isolated line with no known blends, making it adequate for a flare spectral line profile analysis. The atomic database CHIANTI \citep{1997A&AS..125..149D,2012ApJ...744...99L} line list provides an emission temperature of $\log_{10}T$=6.8 ($T$=6.3 MK), but this is calculated with a flare differential emission measure (DEM). Other recent papers state a lower temperature of $\log_{10}T$=6.4 \cite[e.g.][]{2011ApJ...740...70M,2013ApJ...767...83G}, giving a temperature of $T$=2.5 MK. The contribution function $G(T)$ (using CHIANTI) for Fe XVI 262.9760~\AA\, is shown in Figure \ref{fig_goft}, and the peak of $G(T)$ lies close to $\log_{10}T$=6.5 ($T$=3.2 MK) (for both the latest coronal and photospheric abundances). We can also see from Figure \ref{fig_goft} that the $G(T)$ is relatively flat across the peak with temperatures from 2 to 4 MK having a $G(T)$ value larger than half of the peak value.

	The line of Fe XXIII ($263.7657~$\AA) is also free of blending and not closely surrounded by strong spectral lines. Fe XXIII is formed at a high temperature of $\log_{10}T$=7.2. However, the Fe XXIII line profile is often too weak for a confident study, particularly in regions away from the coronal loop-top source. Also, many of the Fe XXIII profiles contain large moving components that complicate the analysis. Hence, Fe XXIII was rejected for this study. However, the study of Fe XXIII might be suitable for other flares. Fe XV ($284.1630~$\AA) might also be suitable for future line profile studies (this line was used by \citet[][published but not yet peer-reviewed]{2013arXiv1305.2939L} for non-flaring coronal observations). Fe XV is a strong line that is formed at $\log_{10}T$=6.4 (close to the peak formation $T$ of Fe XVI). However, for flare SOL2013-05-15T01:45, many of the Fe XV lines contained moving components and we decided that Fe XV was also not suitable for analysis in our chosen regions. Also, there are two weak lines close to Fe XV at 284.1471~\AA~ and 284.0250~\AA, that could pose a problem for a line profile analysis, depending on the conditions and line strengths.

\begin{figure}[ht]
	\centering
	\vspace{-1 cm}
	%\hspace{1 cm}
	\includegraphics[width=0.55\textwidth]{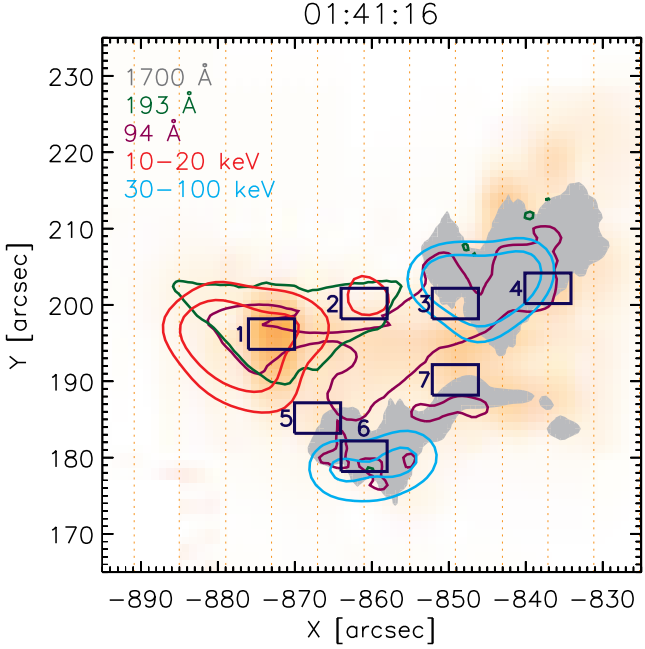}
	\vspace{-0.8 cm}
	\caption{Background EIS integrated intensity raster images for Fe XVI at 262.9760~\AA. The features of the flare are displayed using AIA 1700 \AA\, (grey), 193 \AA\, (green) and {\em RHESSI\,} 10-20 keV (red) and 30-100 keV (blue) contours, for the EIS raster start time of 01:41:16 UT, close to the peak of the flare. Seven regions of study are chosen and they are denoted on the figure as the rectangles 1 to 7. The spatially integrated spectral properties of Fe XVI within each rectangular region are studied.}
	\label{fig3}
	\end{figure}

	In this paper, which focuses on establishing the feasibility of our analysis technique, we examine the Fe XVI line profiles thoroughly at one time interval only, during the EIS raster with a start time of 01:41:16 UT. This raster covers the times of peak SXRs and HXRs. Figure \ref{fig3} shows an EIS integrated intensity raster image for Fe XVI, at this time. The orange dashed lines denote the X position of the EIS slit centre position at $5''.99$ intervals. AIA and {\em RHESSI\,} contours are displayed, showing the X-ray loop-top source, UV ribbons and HXR footpoints in relation to the EIS intensity image. The image shows the positions of seven regions of study covering different areas of the flare: the loop-top (region 1), loop leg (regions 2 and 5), ribbon locations with HXR footpoints (regions 3, 4, and 6), and a ribbon location without HXR footpoint emission (region 7). Each chosen region has dimensions of $X=5''.99$ ($2''$ slit located at the centre of the bin) and $Y=4''$, and we study the spatially integrated emission in each region. The natural binning of the EIS observation in $Y$ is $1''$, but we create $4''$ bins to increase the line profile intensity and reduce the error. Figure \ref{fig1} depicts the temporal changes in the Fe XVI integrated intensity in the regions 1 to 7. There is a large rise in Fe XVI integrated intensity before and during the peak flare times. The peak in Fe XVI appears after the peak in SXR emission in all regions (which might suggest it is more abundant in certain regions as the plasma begins to cool). Importantly, it is present in all regions during the flare times but it has a larger integrated intensity in the loop leg and coronal regions (1, 2, and 5). The plotted integrated intensity is found from single Gaussian fitting that is discussed in Section \ref{profiles}. The integrated intensity of Fe XVI increases by about an order of magnitude or more during the flare times, in all regions, showing that its formation is greatly influenced by onset of the flare. We assume that {\em RHESSI\,} and AIA are aligned without any adjustments required. From image comparison (by eye), this is a good assumption (see Figure \ref{fig3}). AIA and EIS are initially aligned using the Solar Software (SSW) routine eis\_aia\_offsets.pro. This routine aligns the two instruments by co-aligning EIS slot images with AIA images, at a given time, from tabulated pointing information. From the eis\_aia\_offsets.pro documentation, we assume there is an error of $5''$ in $Y$, which is adequate for the study. Any further alignment is performed by eye.

	The EIS spectroscopic observations are dominated by the presence of a broad, Gaussian instrumental profile. As discussed in EIS software note 7 \footnote{http://hesperia.gsfc.nasa.gov/ssw/hinode/eis/doc/eis\_notes/\newline07\_LINE\_WIDTH/eis\_swnote\_07.pdf}, the broadening is not constant but varies with CCD Y pixel. For the regions we study (1 to 7), the changes in instrumental broadening are small, with a constant value of $W_{inst}=0.067\,$\AA\,, where $W_{inst}$ is a Gaussian full width at half maximum (FWHM). This is very broad and it accounts for a large proportion of the observed line profile. For Fe XVI, the expected isothermal broadening due to an underlying Maxwellian distribution of ions at a temperature of $\log_{10}{T}=6.4$ is only $W_{th}=2\sqrt{\ln{2}}\sqrt{2k_{B}T/M}=0.039\,$\AA\,, where $k_{B}$ is the Boltzmann constant and $M$ is the mass of an iron ion. Thus, in the absence of other sources of line broadening and assuming a total Gaussian line shape, the total observed Fe XVI line FWHM should be approximately $W=\sqrt{W_{th}^2 + W_{inst}^2}=0.078\,$\AA\,. Before its launch, the instrumental profile of EIS was laboratory tested (\citet{2006ApOpt..45.8674K}) using a limited set of line observations. Although the line profiles were not analysed rigorously, it was concluded that the instrumental profile was adequately fitted with a Gaussian profile, and this is the assumption made in this paper. This is discussed further in Section \ref{errors}.

	\section{Investigating the line profiles of Fe XVI}\label{profiles}
	To investigate the flaring Fe XVI 262.9760~\AA\, line profiles, we perform three studies. The main study focuses on the EIS raster starting at 01:41:16 UT. However, we initially investigate the Fe XVI line profiles at all flare times ($\sim$01:25 UT to 02:24 UT), and in all regions, using the automatic EIS Gaussian fitting software. After, we study the line profiles during the 01:41:16 UT interval using a higher moments analysis and line fitting. During the line fitting, a convolved kappa-Gaussian line profile (generalised Voigt) is compared with a single Gaussian line profile.
	\subsection{Gaussian line fitting of Fe XVI line profiles}\label{initial_fit}
	Initially, all the Fe XVI line profiles at all flare times are studied using the SSW routine eis\_auto\_fit.pro with both wavelength range restrictions and a line template containing initial estimates of one or two Gaussian lines, that can account for the presence of possible moving components. We initially perform this analysis to find any Fe XVI line profiles suitable for a more thorough study (i.e. strong lines without moving components). Using this routine, Gaussians are fitted to each spectral line of interest using the fitting function mpfit.pro. eis\_auto\_fit.pro automatically corrects for the instrumental effects of slit tilt and orbital variation, that act to shift the wavelength of the line. The Gaussian moments are easily extracted using the routine eis\_get\_fitdata.pro, which provides the integrated line intensity, centroid position and the FWHM. We found that the line profiles of Fe XVI can be adequately fitted with a single Gaussian line profile. However, closer inspection and examination of fit reduced $\chi^{2}$ values reveals that a double Gaussian component fit is often a better choice. At early times, before and during the peaks in HXRs and SXRs, and in regions 2, 3, 4, 5, and 6, a second Gaussian component is often required to account for the presence of a blue-shifted component, travelling towards the observer along the line of sight, which is common during an explosive event such as a flare. However, for other line profiles at different times, often a small secondary Gaussian component tries to compensate for one side of the additional `wings' that appear either side of the main stationary component, suggesting that many of the lines have symmetrical wing broadening that cannot be accounted for by a Gaussian line profile. During our chosen time interval starting at 01:41:16 UT, a directed moving component can be observed in region 6 (southern HXR footpoint region) only. 

	\subsection{A moments analysis of Fe XVI line profiles}\label{fe_moms}
	Using the results of the initial analysis, suitable Fe XVI line profiles at the EIS raster start time of 01:41:16 UT are chosen for further analysis. Firstly, we perform a higher moments analysis using the third (skewness) and fourth (kurtosis) normalised moments. The skewness describes the symmetry of the line, and this is useful for inferring the presence of small components of directed motion. A symmetric line distribution such as a Gaussian will have a skewness equal to 0. The kurtosis describes how the line shape moves away from that of a Gaussian, which has a kurtosis value of 3. The distribution-normalised skewness $S$ and the kurtosis $K$ are calculated for any observed line intensity $I(\lambda)$ [ergs/cm$^{2}$/s/sr/\AA] via
	\begin{equation}\label{skew}
	S=\frac{1}{\sigma^{3}}\frac{\int_{\lambda} I(\lambda) (\lambda-\lambda_{0})^{3} d\lambda}{\int_{\lambda} I(\lambda) d\lambda}
	\end{equation}
	and
	\begin{equation}\label{kurt}
	K=\frac{1}{\sigma^{4}}\frac{\int_{\lambda} I(\lambda)(\lambda-\lambda_{0})^{4} d\lambda}{\int_{\lambda} I(\lambda) d\lambda}
	\end{equation}
	for wavelength $\lambda$ and line centroid $\lambda_{0}$. Both $S$ and $K$ are weighted by $\sigma^{2}$, the second moment (the variance) of the line given by 
	\begin{equation}\label{sigma}
	\sigma^{2}=\frac{\int_{\lambda} I(\lambda)(\lambda-\lambda_{0})^{2} d\lambda}{\int_{\lambda} I(\lambda) d\lambda}.
	\end{equation}
	A sensible range of $\lambda$ values containing the line profiles is chosen ($\lambda_{0}\pm0.2~$\AA), and a background level (assumed constant) found from the initial Gaussian fitting in each region, is removed before evaluating the line profile skewness and kurtosis. The skewness and kurtosis are then found using Equations \ref{skew} and \ref{kurt}. Any variation in wavelength due to the instrumental effects is also accounted for by using the offset provided within the eis\_auto\_fit.pro fit structure. However, the offset will not change the overall shape of the line profile and the analysis. The skewness and kurtosis values for each Fe XVI line are shown in Table \ref{tb1}. The total line width (calculated as a Gaussian FWHM$=2\sqrt{2\ln2\sigma^{2}}$, where $\sigma^{2}$ is the variance), including the instrumental broadening, is also displayed in Table \ref{tb1}. Calculating the total line width as a `Gaussian FWHM' allows comparison with a Gaussian, since the `FWHMs' will only be equivalent when the line shape is truly Gaussian and $\sigma^{2}=\sigma_{G}^{2}$, where $\sigma_{G}^{2}$ is the variance of a Gaussian distribution. From the moments analysis, the total line widths have `FWHM' values between $0.09$~\AA~ and $0.11$~\AA. For any further analysis, we only choose lines with an absolute value of skewness, $|S|\le0.1$. As expected from the initial analysis, the line profile in region 6 has a larger $|S|$ than $0.1$ and is not further analysed. This value of skewness was chosen after performing an analysis of modelled lines with different levels of skewness. This is discussed further in Section \ref{errors}. Table \ref{tb1} shows that the kurtosis values are between 3.2 and 3.5, slightly higher than the Gaussian value of 3. The kurtosis values are suggestive that the line profiles deviate from a Gaussian. If we are measuring the kurtosis of physical line profiles ``gaussianised'' by a broad instrumental profile, we would expect the kurtosis values of the physical line profiles to be higher than measured. There does not seem to be any obvious change in kurtosis related to the different flare regions. The skewness and an uncertainty associated with the measured kurtosis will be discussed in Section \ref{errors}.

\begin{table*}[t]
\begin{center}
     \caption{The 2nd, 3rd and 4th moments of the spectral line profiles between $\lambda_{0}\pm0.2~$\AA. We display the total $\sigma^{2}$, i.e. including the instrumental profile.}
    \begin{tabular}{c c c c}
    \hline\hline
     & & Moments &  \\ \hline
     Region & $2\sqrt{2\ln{2}\sigma^{2}}$~(\AA) & Skewness & Kurtosis\\ \hline
    1 &0.09 & 0.00 & 3.2 \\ \hline
    2 &0.09 & 0.05 & 3.3 \\ \hline
    3 &0.10 & -0.09 & 3.3 \\ \hline
    4 &0.09 & -0.06 & 3.3 \\ \hline
    5 &0.09 & -0.05 & 3.2 \\ \hline
    6 &0.13 & -0.24 & 3.2 \\ \hline
    7 &0.1 & -0.08 & 3.5 \\ \hline
	\end{tabular}
    \label{tb1}
    \end{center}
\end{table*}

\begin{table*}[t]
\begin{center}
     \caption{Line fitting parameters for each region. The error for each $\sigma$ is not shown as it is small ($\sim10^{-3}-10^{-4}$~\AA). The line is fitted across a wavelength range of $\lambda_{0}\pm0.25~$\AA. $\sigma_{G}$ and $\sigma_{\kappa}$ (kappa fit) are shown with the instrumental broadening included. KG=kappa-Gaussian fit.}
    \begin{tabular}{c c c c c c c c c}
    \hline\hline
     &Gaussian& & Kappa&  & & KG &  &  \\ \hline %&  \\ \hline
     Region & $\chi_{G}^{2}$ & $2\sqrt{2\ln{2}\sigma_{G}^{2}}$~(\AA) & $\chi_{\kappa}^{2}$ &$2\sqrt{2\ln{2}\sigma_{\kappa}^{2}}$~(\AA)&$\kappa$& $\chi_{\kappa G}^{2}$ & $2\sqrt{2\ln{2}\sigma_{\kappa}^{2}}$~(\AA) & $\kappa$ \\ \hline%& Equilibrium $T$ (MK) \\ \hline
    1 & 4.1 & 0.09 & 0.6 & 0.09 & 10.4$\pm$1.3 & 0.7 & 0.05 & 3.3$\pm$0.4 \\ \hline%& 4.1$\pm$0.2 \\ \hline
    2 & 1.9 & 0.09 & 0.5 & 0.08 & 7.2$\pm$1.4 & 0.4 & 0.04 & 2.0$\pm$0.3 \\ \hline%& 2.1$\pm$0.4 \\ \hline
    3 & 3.4 & 0.1 & 1.3 & 0.09 & 7.2$\pm$1.2 & 1.2 & 0.05 & 2.6$\pm$0.4 \\ \hline%& 4.5$\pm$0.4 \\ \hline
    4 & 3.9 & 0.09 & 1.2 & 0.09 & 8.0$\pm$1.2 & 1.3 & 0.05 & 2.9$\pm$0.4 \\ \hline%& 4.2$\pm$0.3 \\ \hline
    5 & 1.7 & 0.09 & 0.4 & 0.08 & 6.2$\pm$1.3 & 0.5 & 0.04 & 2.1$\pm$0.4 \\ \hline%&2.7$\pm$0.5 \\ \hline
    6 & - & - & - & - & - & - & - & - \\ \hline %& - \\ \hline
    7 & 5.3 & 0.1 & 1.2 & 0.09 & 5.7$\pm$0.7 & 0.9 & 0.05 & 2.1$\pm$0.2 \\ \hline%& 3.6$\pm$0.3 \\ \hline
	\end{tabular}
    \label{tb2}
    \end{center}
\end{table*}

	\subsection{Line fitting}
	At the very least, we know that the line profile is a combination of two functions: (1) the instrumental profile and (2) the physical profile produced by the motion of the ions.  In order to account for the possibility of non-Gaussian physical line profiles, we employ the more general kappa line profile, that takes the form,
	\begin{equation}\label{Ikappa}
	I(\lambda)=I_{0}\left(1+\frac{(\lambda-\lambda_{0})^{2}}{2\sigma_{\kappa}^{2}\kappa}\right)^{-\kappa}
	\end{equation} 
		for amplitude $I_{0}$ and $\sigma_{\kappa}$, a characteristic width\footnote{We note that $\sigma_{\kappa}^{2}$ in a kappa distribution of this form is not equal to the actual second moment (variance) of the line, but it is related to it by $\sigma^{2}=\sigma_{\kappa}^{2}/(1-3/2\kappa)$ \citep{2015JGRA..120.1607L}. Hence the distribution variance can be easily found from a simple change of variables.} Small values of the index $\kappa$ can produce lines that are more peaked with broader wings since the line profile is produced by a velocity distribution out of thermal equilibrium with a greater fraction of higher velocity particles. Equation \ref{Ikappa} tends to a Gaussian line profile as $\kappa\rightarrow\infty$. From a real-life observational perspective, the line profiles will be indistinguishable from a Gaussian if $\kappa \geqslant 20$. In the high $\kappa$ index limit, the characteristic width $\sigma_{\kappa}^{2}$ has the same meaning as $\sigma_{G}^{2}$, the variance of a Gaussian line profile. The implications of a kappa line profile instead of a Gaussian are discussed further in Section \ref{discuss}. 

	A total observed line profile $W$ can then be written as a convolution of a Gaussian line profile $\mathcal{G}$ and a kappa line profile $\mathcal{K}$,
	\begin{equation}\label{gen_voigt}
	\begin{split}
	\mathcal{W}(\lambda;\sigma_{I},\sigma_{\kappa},\kappa)& = \mathcal{G}*\mathcal{K}\\
	& =\int_{-\infty}^{\infty}
	\exp{\left(-\frac{\lambda^{'2}}{2\sigma_{I}^{2}}\right)}\left(1+\frac{(\lambda-\lambda^{'})^{2}}{2\sigma_{\kappa}^{2}\kappa}\right)^{-\kappa}d\lambda^{'}
	\end{split}
	\end{equation}
	where the integrated intensity here is normalised to $1$ and each function is conveniently centred at $\lambda=0$. $\mathcal{W}$ is similar to a Voigt function which is a convolution of a Gaussian profile and a Lorentzian profile, except that the $\kappa$ index of a kappa distribution can vary and it is not fixed at a value of $-1$ as for a Lorentzian distribution. In some areas of physics $\mathcal{W}$ may be called a generalised Voigt function and a kappa distribution may be called a generalised Lorentzian. As with the Voigt function, there is no readily available analytic form of Equation \ref{gen_voigt}, but it can be found numerically over a range of observation $[-\lambda,\lambda]$ and fitted to the observed line profiles. 
	
\begin{figure*}
	\centering
	\includegraphics[width=8.5cm,angle=0]{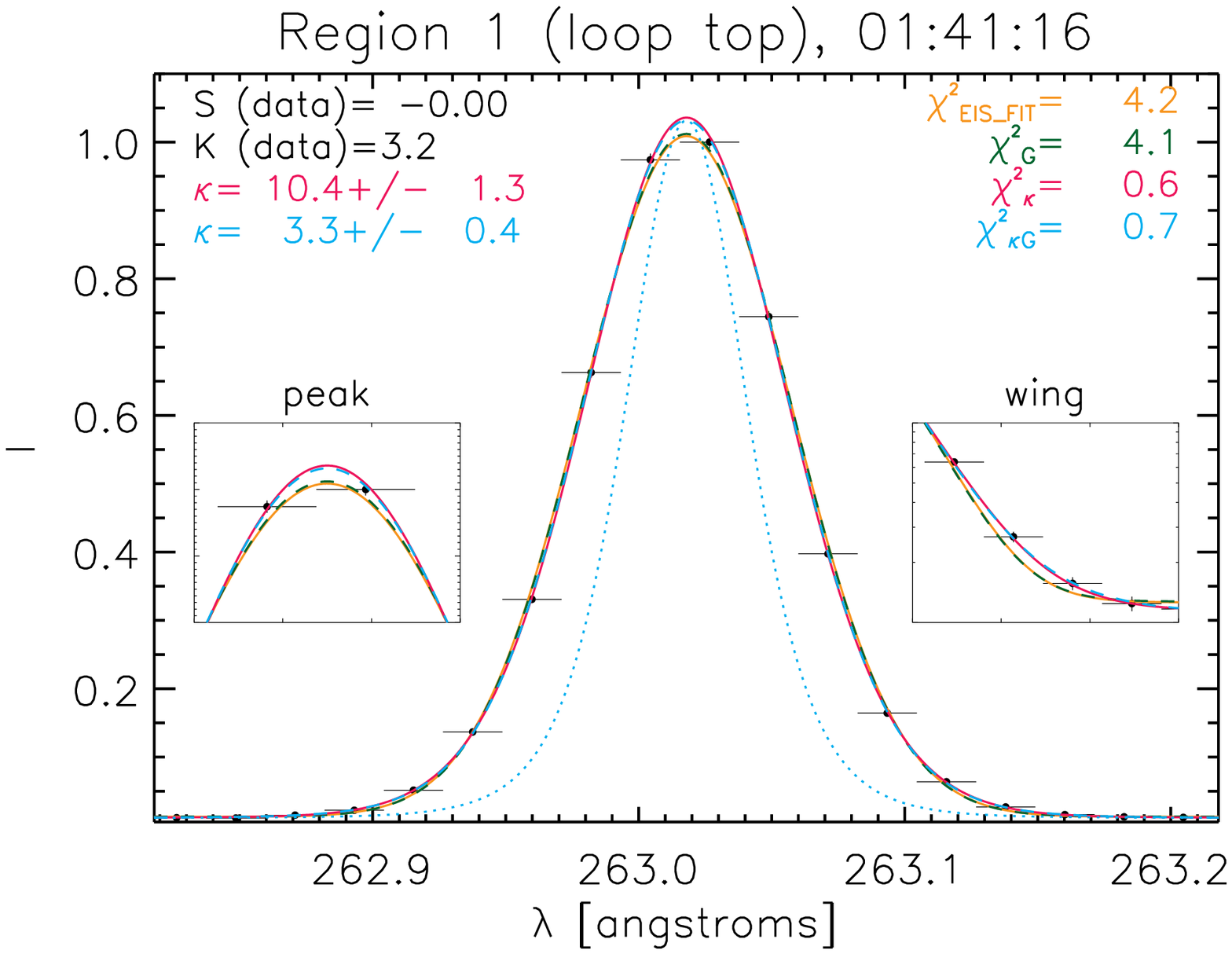}
	\includegraphics[width=8.5cm,angle=0]{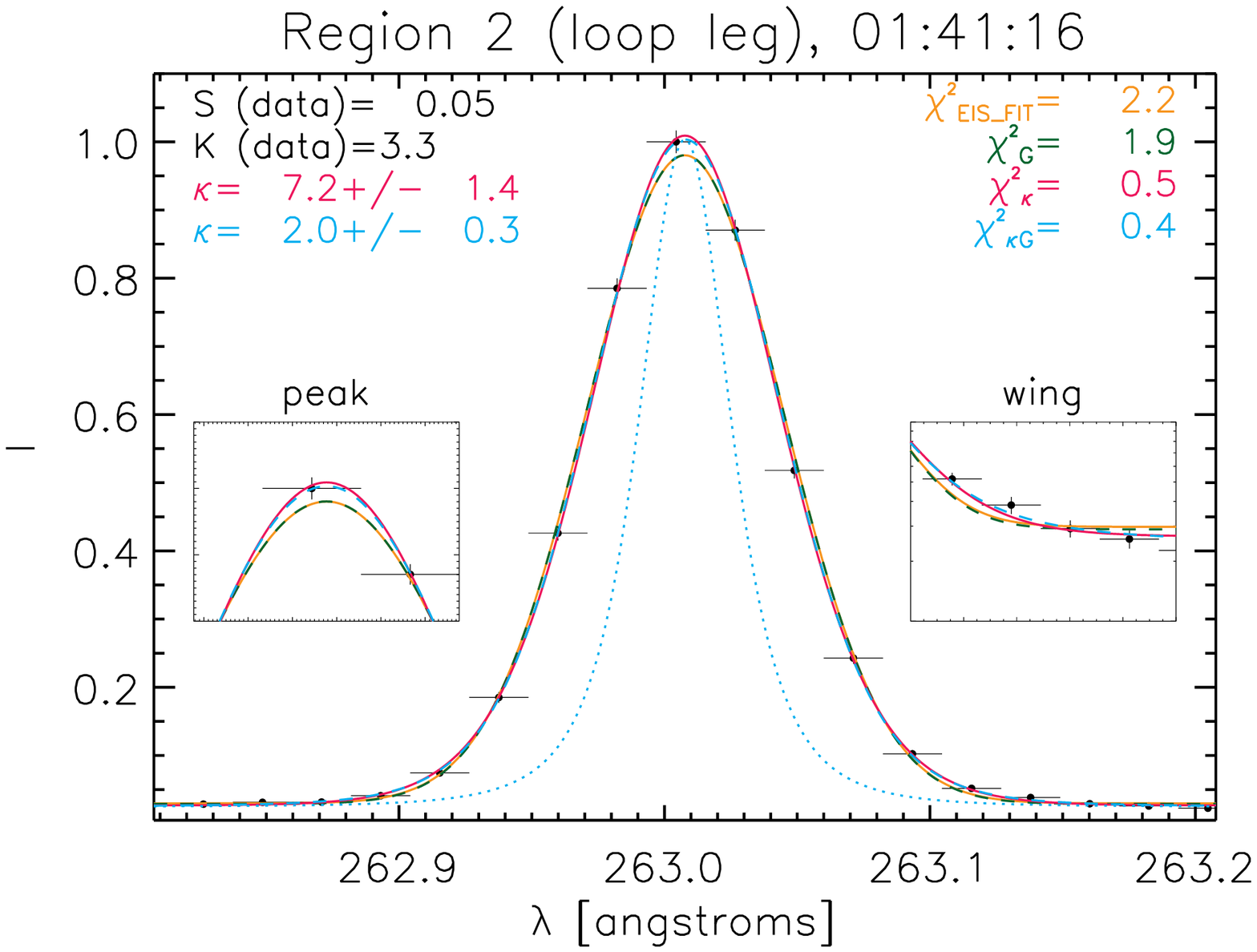}
	\includegraphics[width=8.5cm,angle=0]{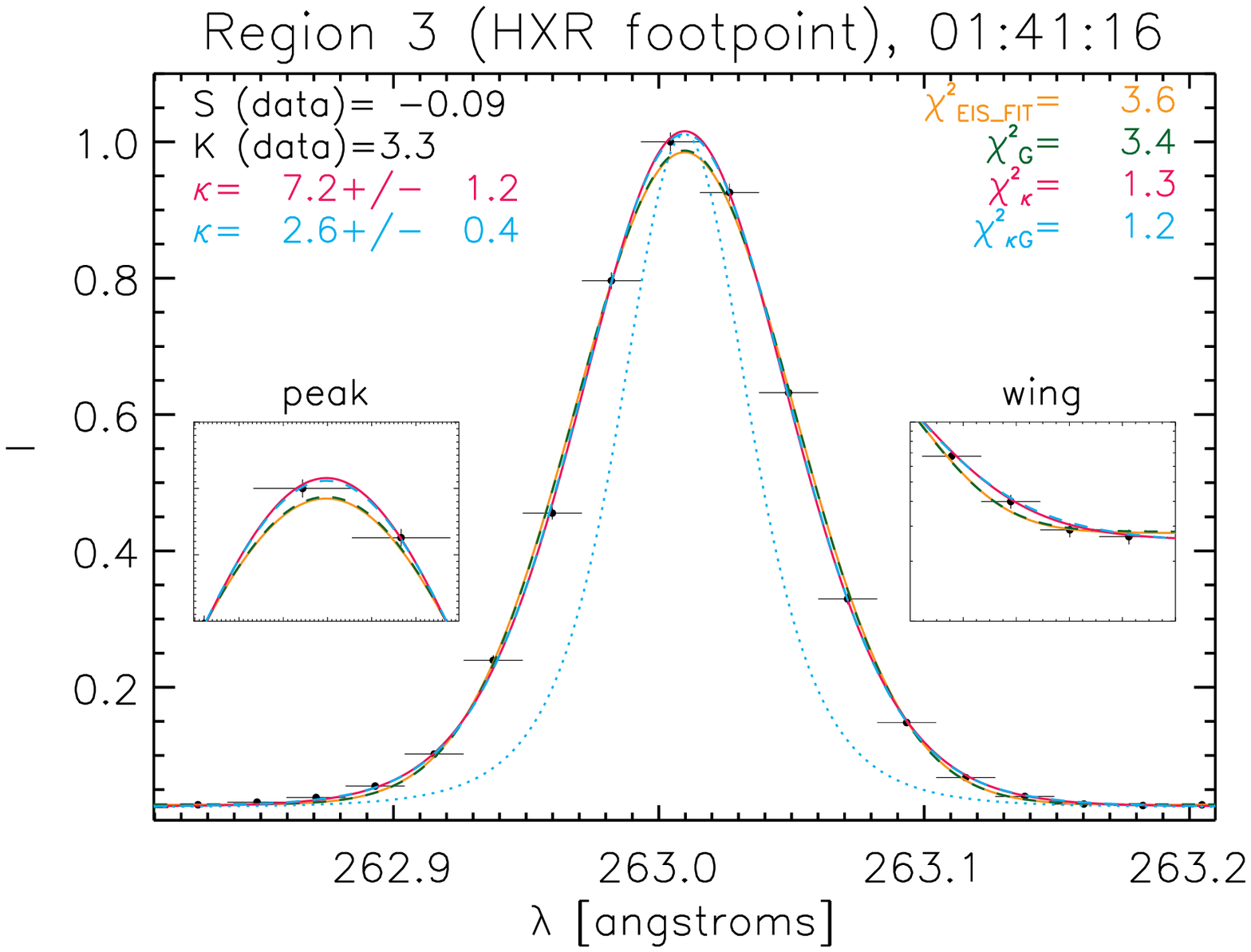}
	\includegraphics[width=8.5cm,angle=0]{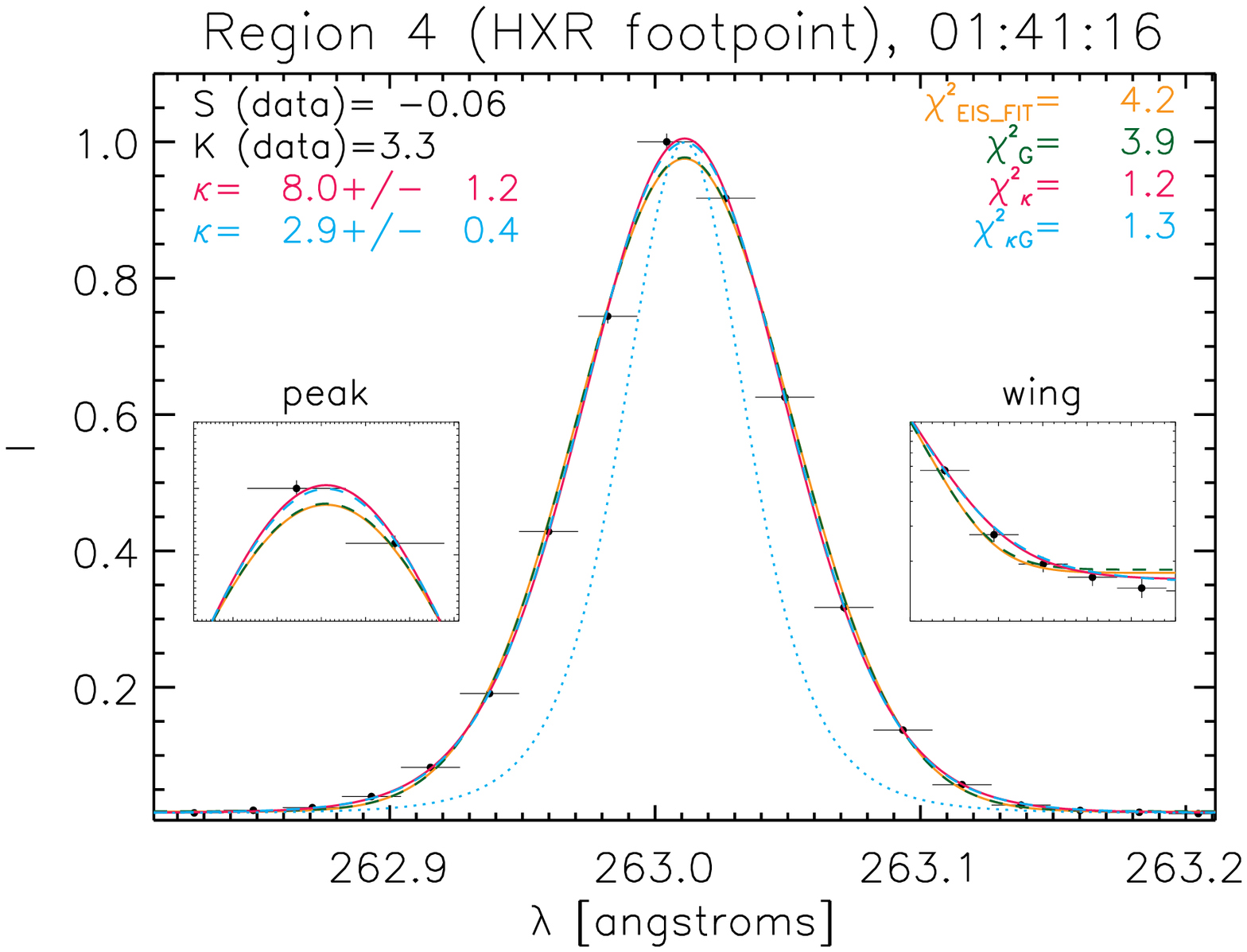}
	\includegraphics[width=8.5cm,angle=0]{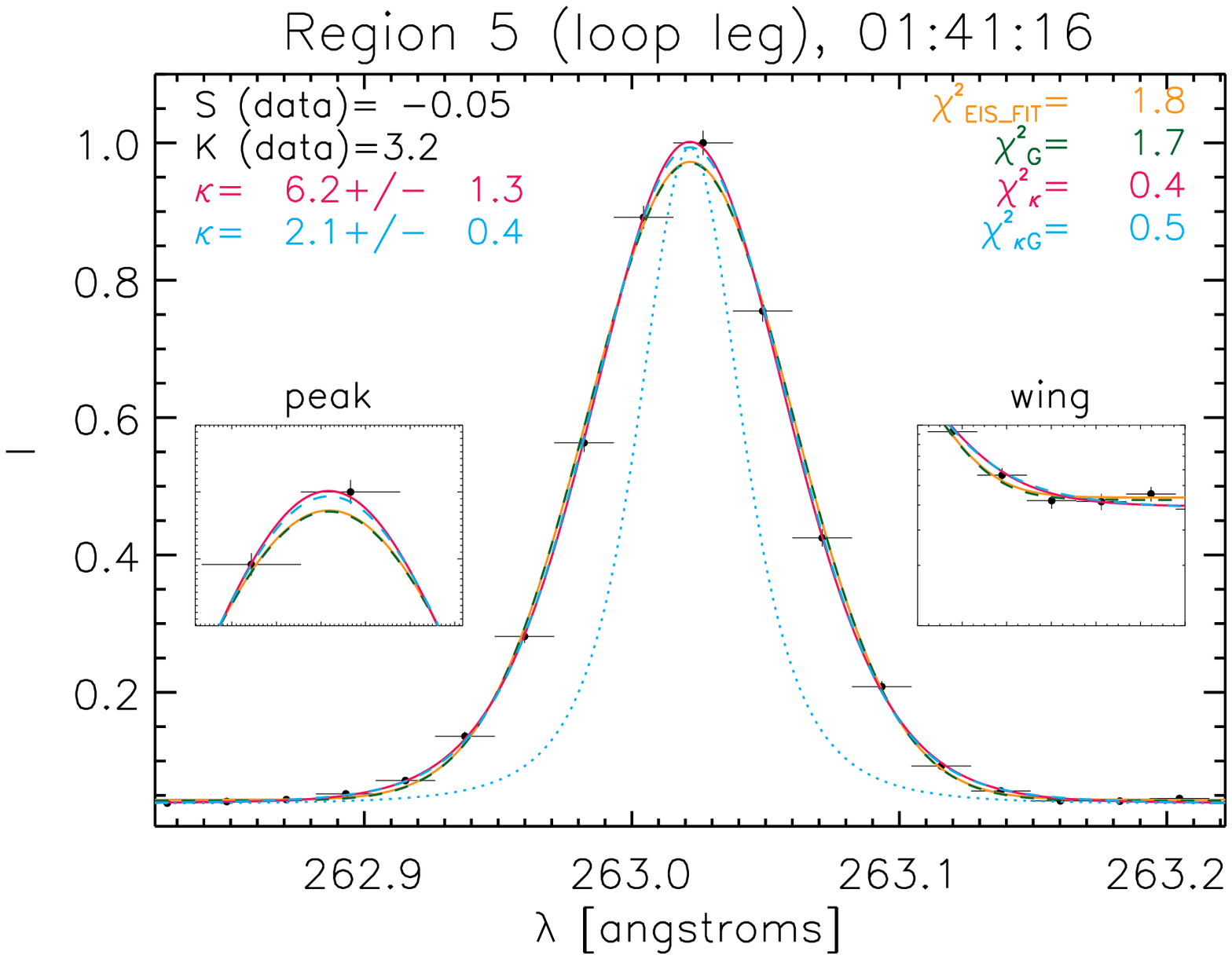}
	\includegraphics[width=8.5cm,angle=0]{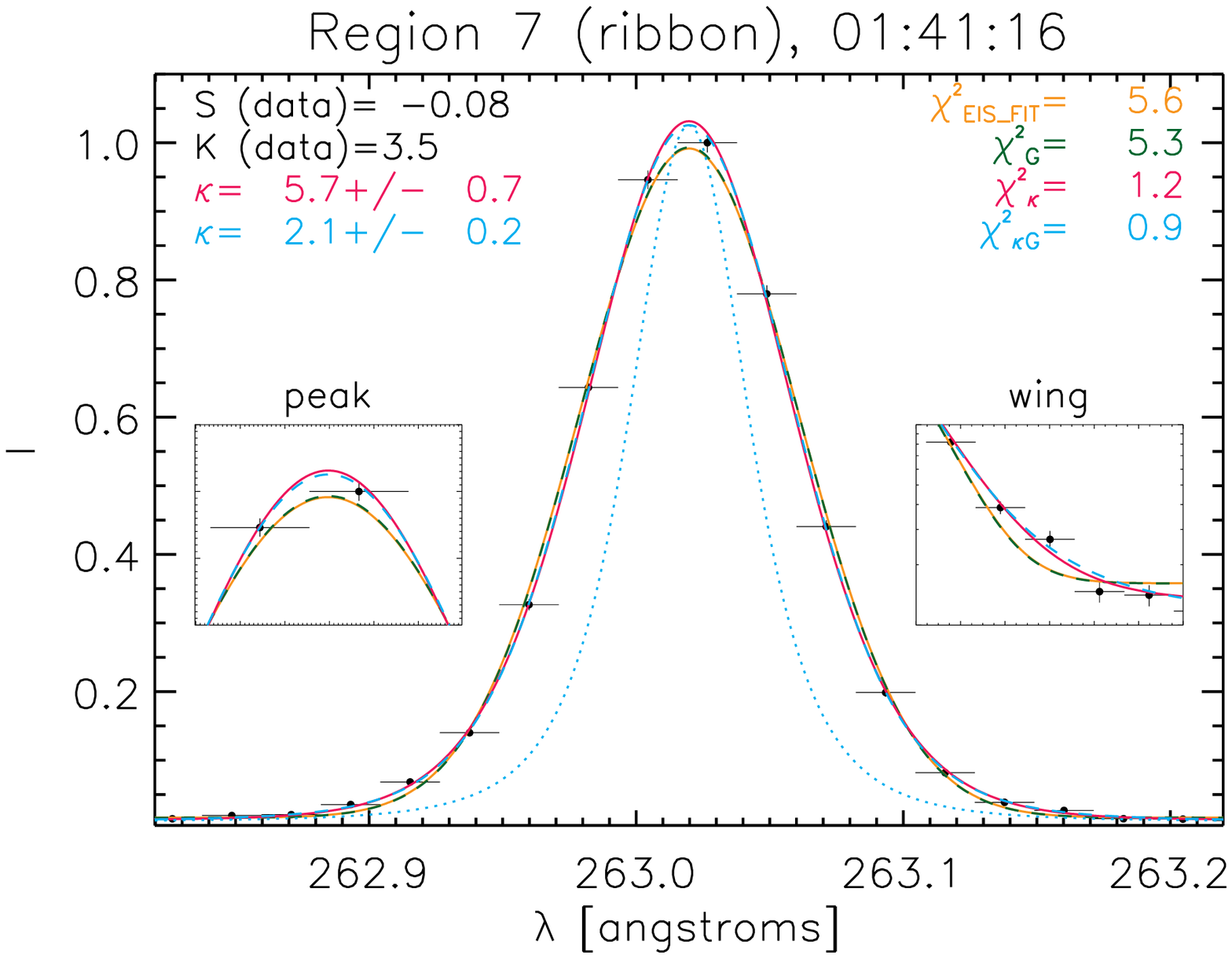}
	\caption{Fe XVI line profiles (total observed profile) for the regions of 1, 2, 3, 4, 5, and 7. The peak of each line is set to 1 by dividing by the maximum value. Many of the lines are ``more peaked'' and have broader wings, which is consistent with a physical line profile closer to a kappa distribution than a Gaussian. Small panels on the main plots show the peaks and wings more clearly. The following fits are shown: orange: Gaussian from eis\_auto\_fit, green: Gaussian fit, pink: kappa fit, blue: convolved kappa-Gaussian fit. The reduced $\chi^{2}$ values for each fit are shown on each panel. The skewness ($S$) and kurtosis ($K$) from the moments analysis, and the $\kappa$ values of each fit, are also shown. The dotted blue lines represent the inferred physical kappa line profiles from the convolved kappa-Gaussian fits.}
	\label{fig10}
	\end{figure*}

	By fitting a convolved $\mathcal{W}$ to the profiles, the kappa line profile fit parameters are directly related to the underlying physical processes, as the instrumental profile is automatically accounted for by the Gaussian. As with the moments analysis, only lines with a small skewness of $|S|\le0.1$ and no obvious secondary (moving) components are fitted. For fitting purposes, Equation \ref{gen_voigt} can be rewritten as a discrete convolution with fit parameters $A$ as,
	\begin{equation}\label{IW}
	\begin{split}
	\mathcal{W}(\lambda) & = \mathcal{G}*\mathcal{K}= A[0]+\\ 
	& A[1]\sum_{\lambda^{'}}\exp{\left(-\frac{(\lambda^{'}-A[2])^{2}}{2\sigma_{I}^{2}}\right)\left(1+\frac{(\lambda-\lambda^{'}-A[2])^{2}}{2A[3]^{2}A[4]}\right)^{-A[4]}}
	\end{split}	
	\end{equation}
	with an added background $A[0]$. We fit convolved kappa-Gaussian profiles, $\mathcal{W}$ to the lines where the fixed FWHM$=2\sqrt{2\ln{2}\sigma_{I}^{2}}=0.067~$\AA represents the Gaussian width of the instrumental profile. The five fit parameters $A$ are all free, and are found via the fitting procedure (using mpcurvefit.pro). For comparison, the lines are also fitted with a single kappa line profile of the form,
	\begin{equation}\label{Ikappa_fit}
	I(\lambda)=B[0]+B[1]\left(1+\frac{(\lambda-B[2])^{2}}{2B[3]^{2}B[4]}\right)^{-B[4]}
	\end{equation}
	with free fit parameters $B$, and a single Gaussian line profile,
	\begin{equation}\label{IGaussian_fit}
	I(\lambda)=C[0]+C[1]\exp\left(-\frac{(\lambda-C[2])^{2}}{2C[3]^{2}}\right).
	\end{equation}
	with free fit parameters $C$. A comparison of the convolved kappa-Gaussians fits, kappa fits, Gaussian fits and single Gaussian fits using the EIS automatic fitting procedure eis\_auto\_fit.pro are shown in Figure \ref{fig10}, for the Fe XVI lines, in all regions apart from 6, which was not suitable for further analysis. Examining the Fe XVI lines shows that, as well as broader wings, many of the lines have a peaked feature that can not be fitted by a single Gaussian distribution. Such line shapes can be better fitted by the extra free $\kappa$ index parameter in the convolved kappa-Gaussian function. The reduced $\chi^{2}$ values from each of the fits are shown. There are three sensible scenarios:
	\begin{enumerate}
	\item $\chi^{2}_{\kappa G}$ low, $\kappa$ index high and $\chi^{2}_{G}$ low $\rightarrow$ $\sim$ Gaussian.
	\item $\chi^{2}_{\kappa G}$ low, $\kappa$ index low and $\chi^{2}_{G}$ low $\rightarrow$ indeterminable.
	\item $\chi^{2}_{\kappa G}$ low, $\kappa$ index low and $\chi^{2}_{G}$ high $\rightarrow$ $\sim$ Kappa.
	\end{enumerate}
	Fits where the $\chi^{2}$ values are within $\le2$ of each other or the $\chi^{2}$ values both lie within $0<\chi^{2}<2$ are deemed indistinguishable. In regions 1, 3, 4, and 7, representing the loop-top, HXR footpoints and ribbon, the $\chi^{2}$ values of the Gaussian fits are more than double that of the kappa-Gaussian convolved fits. The kappa-Gaussian convolved $\chi^{2}$ values in these regions are also close to 1, with values between 0.7 and 1.3, while the Gaussian values vary between 3.4 and 5.3 (Table \ref{tb2}). In regions 2 and 5, the `loop-leg' regions, both the kappa-Gaussian and Gaussian $\chi^{2}$ are close to 1.  Here the kappa-Gaussian convolved $\chi^{2}$ values are 0.4 and 0.5, while the Gaussian values are 1.9 and 1.7, and we cautiously suggest that their form cannot be confidently found from the fitting. From the kappa-Gaussian distribution fits, $\kappa$ index values of 2.0 to 3.3 are found, and these are displayed in Table \ref{tb2}. The uncertainty in each $\kappa$ index value is found from the fit and it is small, less than $1$ for the convolved kappa-Gaussian fits. 

	The characteristic widths found from the kappa-Gaussian fitting are also shown in Table \ref{tb2}. As with the second moment found in Section \ref{fe_moms}, the width is written as a `Gaussian FWHM=$2\sqrt{2\ln{2}\sigma_{\kappa}^{2}}$', for easy comparison with an actual Gaussian FWHM. The values of $2\sqrt{2\ln{2}\sigma_{\kappa}^{2}}$ are between 0.04~\AA~ to 0.05~\AA. The Gaussian FWHMs, after the removal of the instrumental broadening via quadrature, are between 0.06~\AA~to 0.07~\AA. This will be discussed further in Section \ref{discuss}.

	From Figure \ref{fig10}, we also note that the line centroids at 01:41:16 UT are red-shifted to 
	$\sim$263.01~\AA--263.02~\AA, compared to the laboratory wavelength of 262.9760~\AA. However, this does not change the line profile analysis.

	\section{Uncertainties and problems associated with the determination of line shape}\label{errors}
	Before discussing the results, we consider the different sources of uncertainty associated with the line profile analysis. The main sources of uncertainty are related to: (i) the presence of unknown blended lines, (ii) other lines located close by, (iii) small, hard-to-see moving components and line skewness, (iv) the instrumental profile and broadening, (v) the wavelength range across the line used for the analysis, (vi) the instrument spectral pixel size, (vii) a good estimation of the background level and (viii) the error associated with each measurement (Poisson and instrumental uncertainies related to the EIS CCDs). In Section \ref{profiles} the uncertainties associated with the line fitting parameters, particularly the $\kappa$ index, were discussed and are shown in Figure \ref{fig10} and Table \ref{tb2}. 

	For all line fits, the background level was assumed to be constant across the line. This is a valid approximation since the lines are studied over a small wavelength window of only $0.5~$\AA~ at the most. The estimation and the removal of the background will only become a problem if the levels of noise are high (probably larger than $\sim$ 10 \%), as this could produce a large uncertainty in the background value.

	We estimate the uncertainty associated with the kurtosis values found from the moments analysis. In Figure \ref{fig_eu}, either the kurtosis (rows 1-3), or the $\kappa$ index (from line fitting - row 4), is plotted against the known $\kappa$ index values of a modelled kappa line convolved with a Gaussian with FWHM$=0.067$~\AA, producing modelled line profiles with parameters close to the observed line profiles. We show how the determination of the kurtosis is affected by three different sources of uncertainty individually: (1) spectral pixel size, (2) wavelength range and (3) line skewness. We also show how the line skewness can change the $\kappa$ index determined from line fitting. In each panel, only the specified variable is set to that of the observation or EIS instrument value (i.e. $\Delta\lambda=0.022~$\AA or $\lambda_{0}\pm0.2~$\AA, for example), while all other parameters are kept at ideal, hypothetical values, i.e. very small pixel size of $\Delta\lambda=0.00067~$\AA, or a large wavelength range $\lambda_{0}\pm1000~$\AA, or a skewness equal to 0.

	\begin{figure*}
	\centering
	\includegraphics[width=17cm]{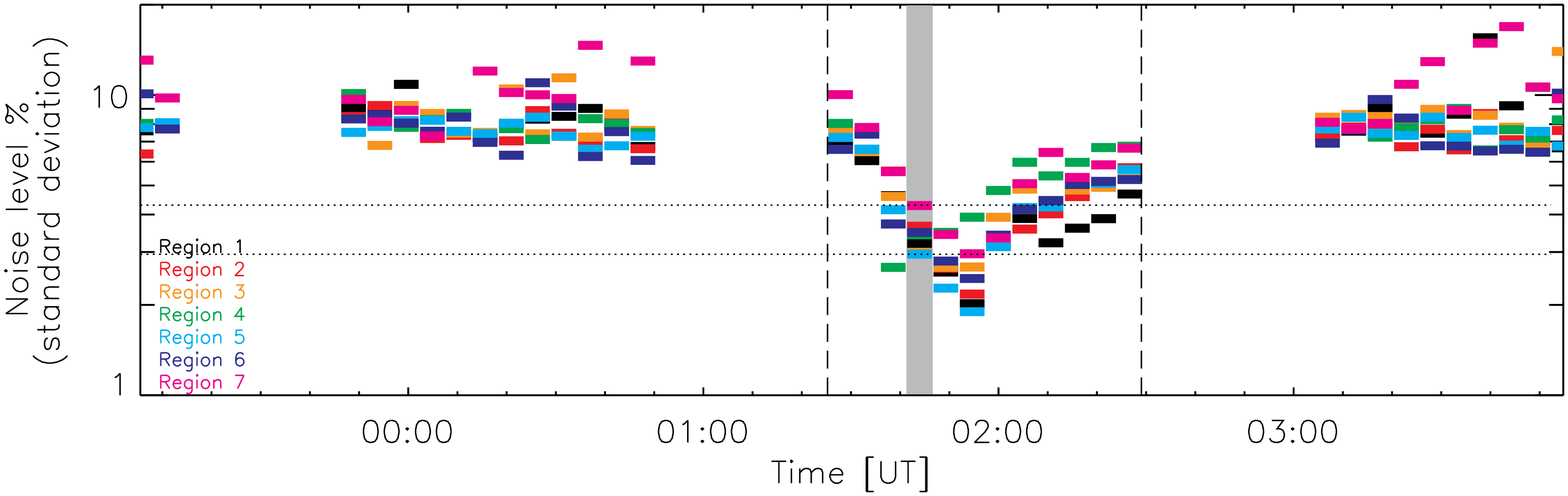}
	\caption{The noise level associated with each Fe XVI spectral line at all times in regions 1 to 7. A 1-sigma Gaussian noise level for each region at each time is calculated using noise(\%)=STD($\epsilon/I$)$\times100$. As expected the noise level falls as the line intensity increases during the flare times of $\sim$01:25 UT to 02:30 UT. The 1-sigma noise level is less than 10 \% during all flare times, and as low as 2 \% during peak flare times. Our studied flare time of 01:41:16 UT is shown by the grey band, with noise levels between $\sim$3-5 \% (shown by the black dotted lines).}
	\label{fig_obs_noise}
	\end{figure*}
	\subsection{Error level associated with each intensity value}
	During the line fitting analysis, only the intensity error was considered. The error value of each intensity measurement $I$ [ergs/s/cm$^{2}$/sr/\AA], for each spectral line, is provided by the EIS software when the data is prepped from level-0 to level-1. As noted in EIS software note 1, the intensity errors are computed assuming photon statistics together with an error estimate for the dark current, and are found when the basic CCD signal or `data numbers' ($DN$) are converted to photons ($P$) and then to intensity units ($I$). The error consists of a combination of photon counting noise (the square root of the photon number) and the 1-sigma error estimated for the dark current, added in quadrature. Once the intensities have been calculated, the overall intensity error is then given by $\epsilon_{I}=\epsilon_{P}I/P$.
	
	For the purposes of investigating uncertainties, and for an estimation of the `noise' level associated with a spectral line profile in a certain region, at a given time, a 1-sigma Gaussian noise level associated with each spectral line is calculated via:
	\begin{equation}
	{\rm Noise(\%)}=100\times{\rm STD}\left(\frac{\epsilon}{I}\right)
	\end{equation}
	where $I$ and $\epsilon$ are the measured intensity and error values respectively and STD denotes the standard deviation. The estimated 1-sigma level of Gaussian noise (\%) in each region is displayed in Figure \ref{fig_obs_noise}, for all regions 1 to 7. During the flare times, the 1-sigma level of noise falls to less than 10\%, with values as low as 2\% during the peak flare times, in all regions, due to the increased Fe XVI intensity. At our studied time (01:41:16 UT), the noise levels vary between $\sim$3-5 \%, denoted by the black dotted lines in Figure \ref{fig_obs_noise}. 

	\subsection{Spectral pixel size}
	EIS has a spectral pixel size of $\Delta\lambda=0.022~$\AA. The effects of changing the spectral pixel size for hypothetical instruments were studied using modelled spectral lines similar to those of Fe XVI observed with EIS. From the first row of Figure \ref{fig_eu}, changes in spectral pixel size from $\Delta\lambda=0.00067~$\AA to that of the EIS value of $\Delta\lambda=0.022~$\AA, should not produce significant differences in the value of kurtosis alone, if all other parameters are kept at the ideal values. The EIS spectral pixel size of $\Delta\lambda=0.022~$\AA has a negligible effect on the line fitting parameters.

	\subsection{Available wavelength range}
\begin{figure}[hpb!]
	\centering
	
	\includegraphics[width=0.8\linewidth]{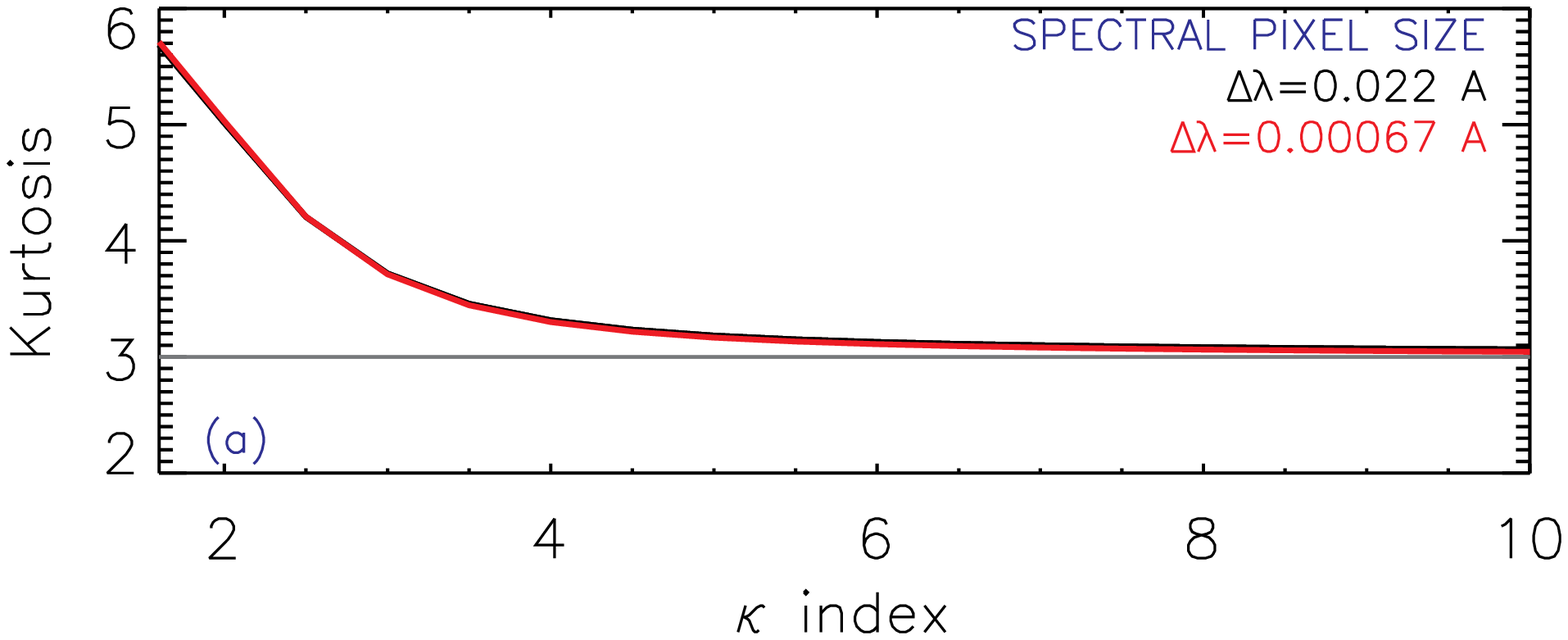}\vspace{0.3cm}
	\includegraphics[width=0.8\linewidth]{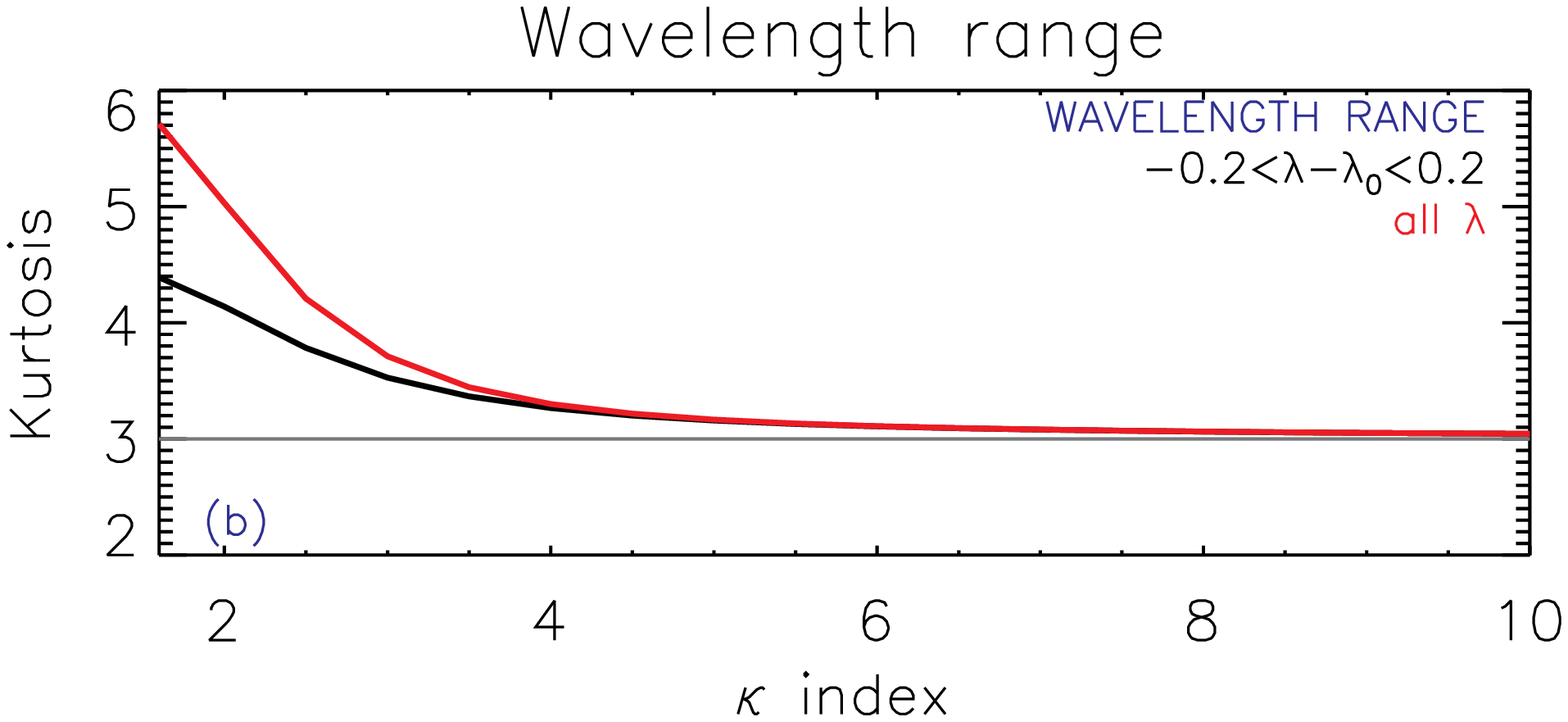}\vspace{0.3cm}
	\includegraphics[width=0.8\linewidth]{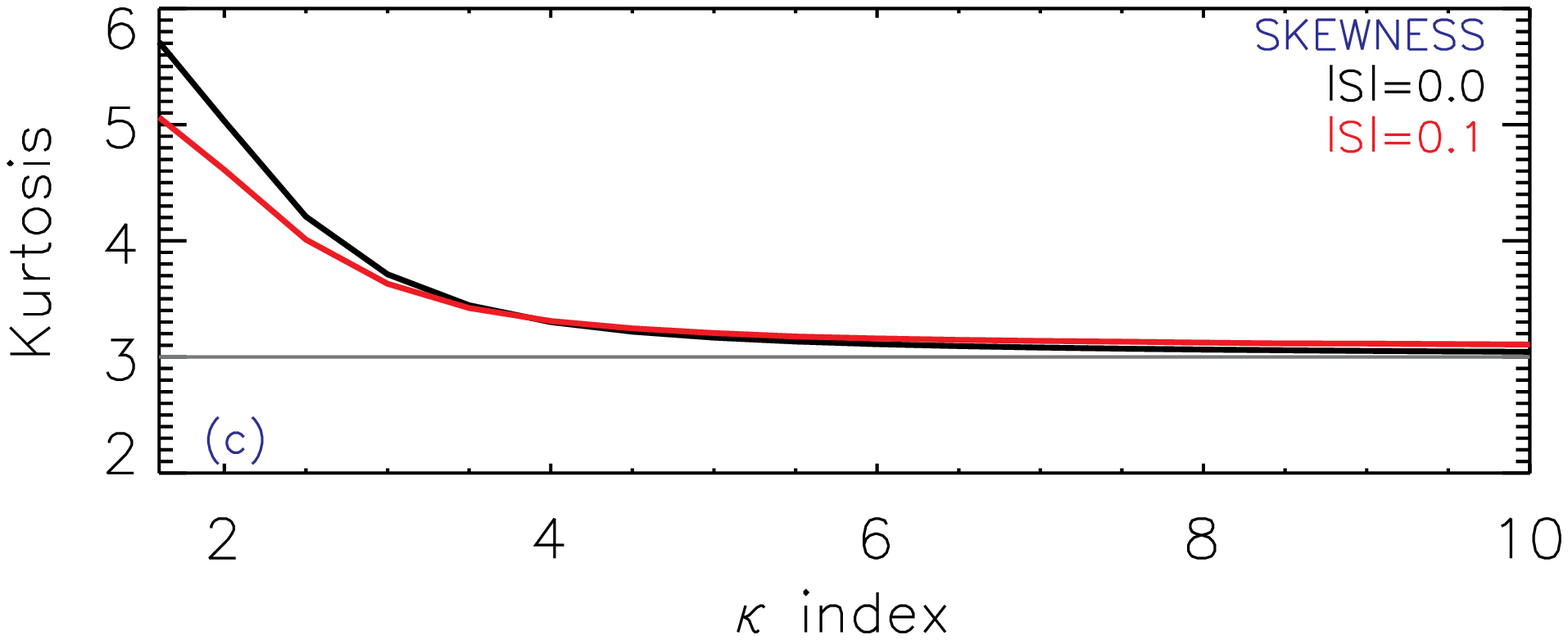}\vspace{0.3cm}
	\includegraphics[width=0.8\linewidth]{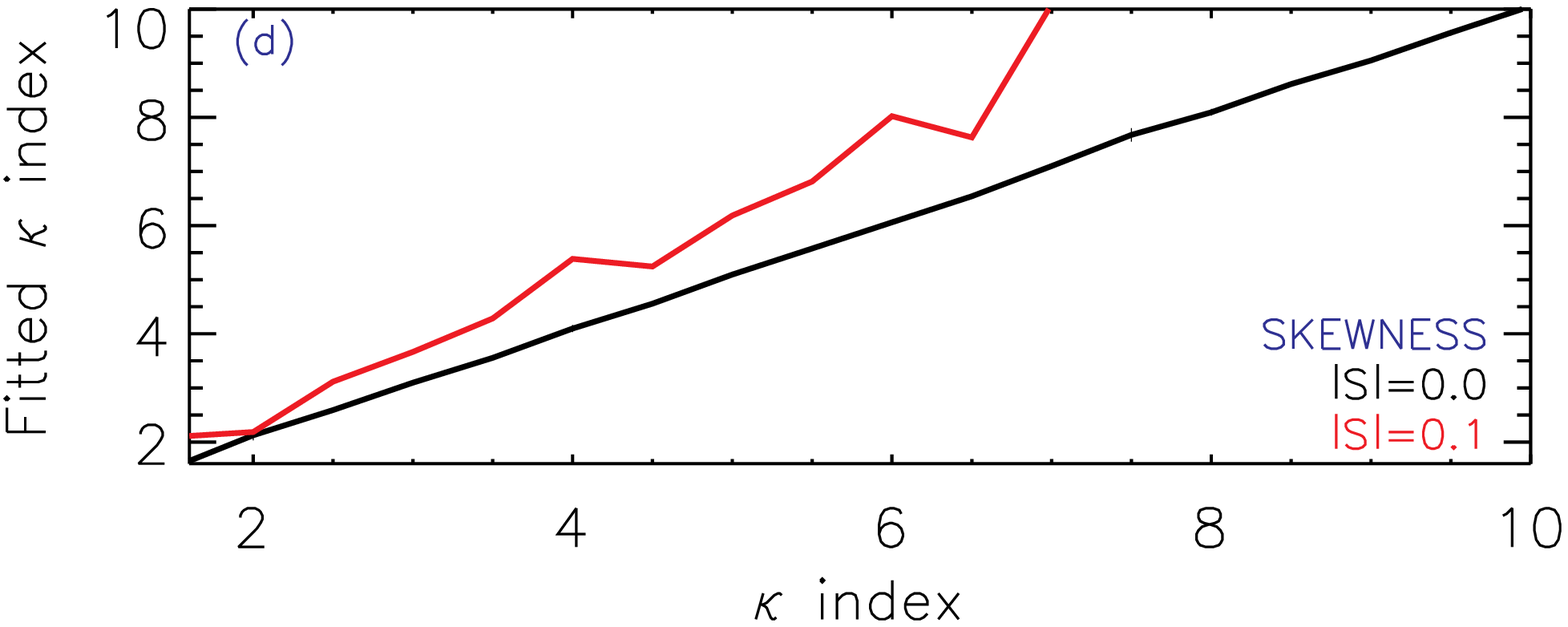}\vspace{0.3cm}
	\includegraphics[width=0.8\linewidth]{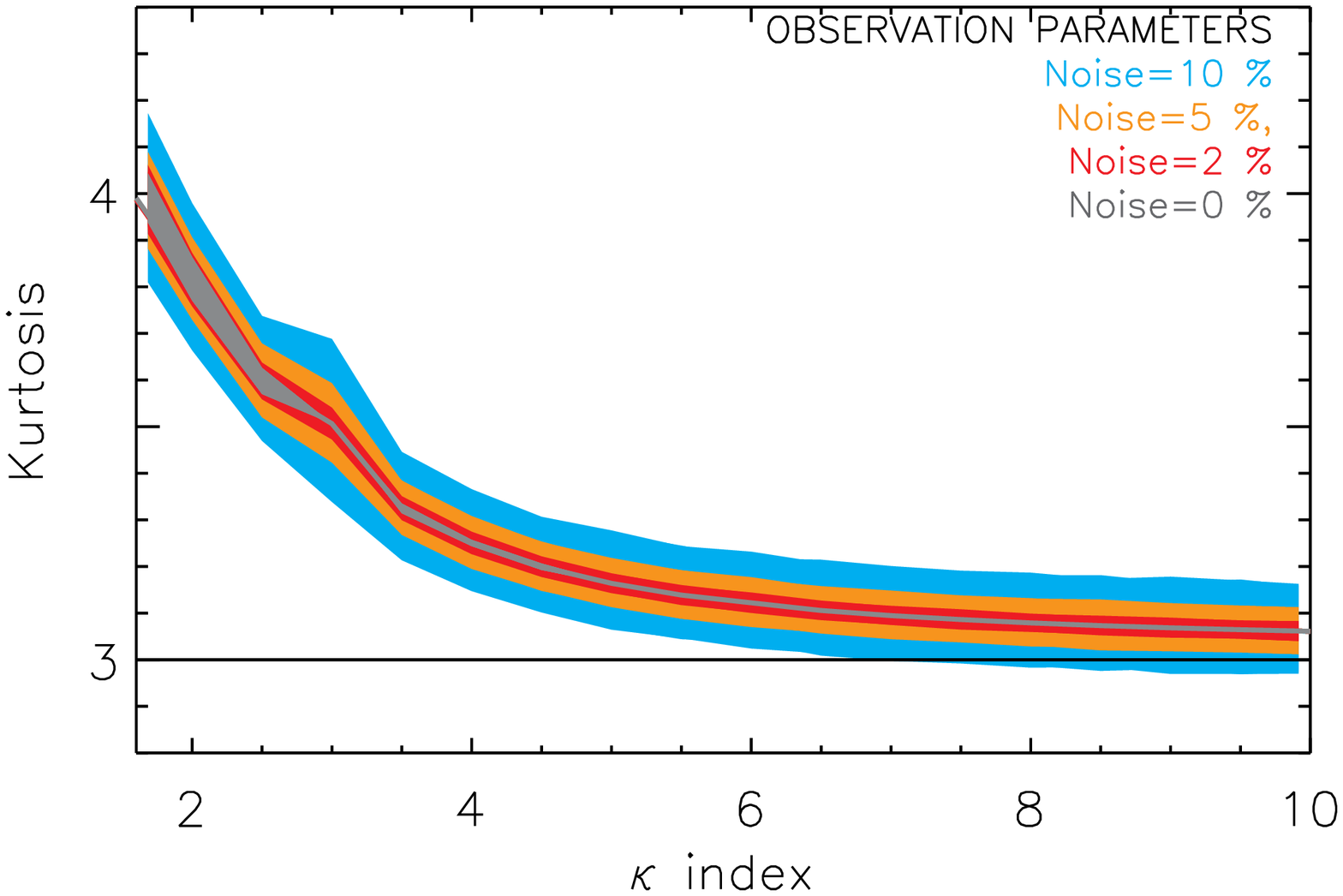}\vspace{-0.5cm}
	\caption{Rows 1-3: The kurtosis value (from a moments analysis) versus the $\kappa$ index of a modelled spectral line, showing differences in the inferred values of kurtosis due to (a) spectral pixel size, (b) chosen wavelength range, and (c) skewness. In each plot, ideal parameters are used and one variable is changed (plot legend) to match the EIS value. Row 4: The $\kappa$ index found from line fitting versus the $\kappa$ index of a modelled spectral line. Line skewness can change the values of $\kappa$ index found from line fitting. Row 5: Kurtosis versus modelled $\kappa$ index, as rows 1-3, but now the collective effects of all sources of uncertainty are shown together using the EIS spectral pixel size ($\Delta\lambda=0.022~$\AA), the observed wavelength range of $\lambda_{0}\pm0.2~$\AA and a skewness equal to 0, for different noise levels of 0 \%, 2 \%, 5, \% and 10 \%.}
	\label{fig_eu}
	\end{figure}

	Often the wavelength range over which the line profile is studied is determined by either the wavelength window of study (as for a EIS observation) or to avoid other lines located close by. The kurtosis of the observed spectral lines was evaluated over $\lambda_{0}\pm0.2~$\AA, to avoid small `lumps and bumps' either side of the line. The line fitting was performed over $\lambda_{0}\pm0.25~$\AA, since such small bumps are not a problem for the line fitting analysis. If all other parameters are ideal, the chosen wavelength range has the biggest influence on the kurtosis value (see Figure \ref{fig_eu} row 2). At low values of $\kappa$, the kurtosis is lower compared to the values found over $\lambda\pm1000~$\AA, falling from $\sim6$ to $4.3$, due to the suppression of the large wings. Therefore, a small wavelength range makes the presence of non-Gaussian line shapes more difficult to determine from a kurtosis analysis. The wavelength range of $\lambda_{0}\pm0.25~$\AA used for line fitting only produces a negligible change in the line fitting parameters, and hence is not shown.

	\subsection{Moving components and skewness}
	The effects of moving components and line skewness were discussed in Section \ref{profiles} and we only studied lines with a small skewness $|S|\le0.1$. This value was chosen from analysing modelled spectral lines using a (relatively large) moving component with a peak intensity $10\;\%$ that of the main component. From this analysis we find that a skewness level of $|S|\le0.1$ should only produce a kurtosis error of $\pm1.0$ at the most, where the $\kappa$ index is very small ($\kappa=1.6$), with the difference decreasing as the $\kappa$ index increases (see Figure \ref{fig_eu} row 3). Also, a line with skewness of $|S|\le0.1$, should be sufficiently symmetrical for analysis. Unlike other uncertainties, we found that the skewness does change the line fitting parameters, especially the $\kappa$ index and this is shown in Figure \ref{fig_eu} row 4, where an inferred $\kappa$ index is plotted against the known $\kappa$ index of the modelled line for $|S|=0$ and $|S|\sim0.1$. We found that the skewness produces a larger difference at high $\kappa$ values ($\sim\pm2$ at $\kappa=6$), but smaller differences at low $\kappa$ values ($<\pm1$ at $\kappa=3$). For the $\kappa$ values inferred from observation, the error should not be larger than $\sim1$. We also note that a higher skewness increases the observed $\kappa$ index value. For kurtosis, larger values of skewness decrease the kurtosis value, and hence the true kurtosis should be larger.
 
	In Figures \ref{fig_eu} row 5, we take all sources of uncertainty into account collectively and instead of assuming ideal parameters, we use only observational/EIS parameters to estimate the overall level of uncertainty associated with the observed kurtosis values, for lines with a physical kappa component convolved with an instrumental Gaussian, for different values of $\kappa$ index. In Figure \ref{fig_eu}, row 5, we set the skewness to 0. We find that the error values for the observed kurtosis values (independent of the skewness) lie somewhere between $\pm0.1$ at the $10 \%$ noise level and $<\pm0.05$ at the $2 \%$ level. The error associated with the line fitting will not vary greatly with noise as long as the errors are accounted for in the fitting process. However, the line shape becomes indeterminable from the fitting process when there are large errors/levels of noise ($\sim>10$~\%).

	There are also a number of instrumental effects that could change the spectral line shape and the results of our analysis. These possible problems are now discussed in turn.

	\subsection{The EIS instrumental profile and broadening}
	As discussed, we assume that the EIS instrumental broadening is completely Gaussian. If not, then it will be partly accounted for by the kappa part of the fitting function. This would act to increase any physical line $\kappa$ index value. Since the instrumental broadening dominates the overall line shape, if the instrumental broadening were highly non-Gaussian then it would be present in every observed spectral line and we would never find lines with a true Gaussian line shape. However, this is difficult to test since the noise level is often too high in quiet Sun regions for a confident analysis (especially for the lines suitable for a profile analysis such as Fe XVI and Fe XXIII). Since the shape of the EIS instrumental broadening was not tested rigorously before launch, it is difficult to discuss further whether the instrumental broadening could be responsible, or at least partly responsible for the non-Gaussian line profiles. This is the largest source of concern for our analysis, and the topic of ongoing work where we are testing whether the instrumental profile could be responsible for the observed non-Gaussian line profiles.

	\subsection{Burn-in effects}
	Another problem could be possible ``burn-in'' effects (as seen for {\it SOHO\;} CDS - for example see \citet{2000OptEn..39.2651T,2010A&A...518A..49D}), whereby the CCD sensitivity in a certain pixel falls over time due to constant exposure to solar radiation. This could cause changes in line shape particularly close to the peak intensity, i.e. a flattening. However, we have been told that the EIS detectors are cleared before every exposure so burn-in should not be a problem. The level of burn-in is adjusted with time (private communication with Louise Harra.) Also, the studied lines show both symmetrical wing broadening and they are more peaked, i.e. the profiles are not consistent with burn-in flattening.

	\subsection{Warm and hot pixels}
	Hot or warm pixels are single pixels that have anomalously high DN values on the EIS CCD. These pixels are marked as `missing' but as we use the ${\rm refill}$ option in eis\_prep.pro, new values for these pixels are found by interpolation of nearby pixels, and such pixels can be found by their corresponding $-100$ values in the intensity error array. Within the seven regions of study, none of the pixels are marked as missing (i.e. $-100)$, so we do not have to worry about the presence of warm pixels and any line profile changes they could produce. However, since the effects of warm pixels on line shape are unknown, we aim to investigate this for further studies.

	\subsection{JPEG compression}
	JPEG compression is used when EIS data is transmitted. During our study, the data was compressed using JPEG75. We are informed that the JPEG compression in most cases should only add a certain level of noise to the data, depending on the level of compression and it should not change the shape of the line. JPEG compression might only pose a problem to line shape when there are very strong gradients in the intensity, near coronal holes for example (``Notes on the compression of EIS spectral data''\footnote{EIS software note 11 \url{http://hesperia.gsfc.nasa.gov/ssw/hinode/eis/doc/eis_notes/11\_JPEG\_COMPRESSION/eis\_swnote\_11.pdf} by Harry Warren} and private communication with Harry Warren). However, since we are only studying small regions in a single flare, this is not an issue for our analysis. We have studied how Gaussian noise changes the line profile and we are confident that this should not pose a problem for our line profile analysis, since we only perform a line profile study when the noise level is less than $\sim10\%$, as discussed. 

	\subsection{Other possible instrumental and data processing effects}
	\citet{2016SoPh..291...55K} describes an issue regarding intensity changes due to bin averaging. We tested the lines after recalculating the intensity of the points using ${\rm icsf.pro}$ from \citet{2016SoPh..291...55K} and we found no appreciable difference in the results. For instance, the kappa-Gaussian $\chi^{2}$ values varied by $\sim0.3$ at the most, but the Gaussian $\chi^{2}$ values were slightly worse and increased by around 1. However, the parameters of the fit such as the $\kappa$ index did not vary greatly, the `new' values are within the errors of the `old' values. Therefore, we are confident that using the intensity values calculated using ${\rm icsf.pro}$ do not change our line fitting results. If anything, it actually improved the results, producing a greater difference between the kappa-Gaussian and Gaussian $\chi^{2}$ values in regions 2 and 5 (`loop-leg').

	\section{Interpretation of the results}\label{discuss}
	During the analysis, we found evidence (taking into account the collective uncertainties) of non-Gaussian Fe XVI line profiles in regions 1, 3, 4, and 7 (covering the coronal loop-top source, northern HXR footpoint and the southern ribbon) of the flare using two independent studies of higher moments (kurtosis) and line fitting. We will now evaluate the possible causes of the observed non-Gaussian line profiles. Ignoring the possibility of a non-Gaussian instrumental profile, the presence of kappa line profiles could be due to the following three physical processes, independently or collectively:
	\begin{enumerate}
	\item An isotropic but accelerated microscopic non-Maxwellian heavy ion velocity distribution.
	\item Turbulent fluctuations of macroscopic plasma parameters: density, velocity or temperature.
	\item Multi-thermal temperature distribution along the line of sight.
	\end{enumerate}
	(3) is beyond the scope of this paper and is the subject of additional ongoing work. Turbulent plasma fluctuations (2) have already been discussed for laboratory plasmas \cite[see for example][]{2004CoPP...44..283M}. In such a scenario, \cite{2004CoPP...44..283M} show that both density and temperature fluctuations do not change the Doppler-broadened spectral line profile, only fluid velocity fluctuations, and this is a valid possible cause of non-Gaussian line profiles. However, in this paper we narrow our discussion to case (1), microscopic non-thermal ion motion.

\subsection{Non-thermal ion motion}
	The `kappa part' of the convolved kappa-Gaussian line profiles we used for line fitting can be converted to a line-of-sight one-dimensional ion velocity distribution via 
	\begin{equation}\label{conversion}
	I(\lambda)d\lambda \propto f(v)dv\rightarrow f(v)\propto I(\lambda)\frac{d\lambda}{dv}=I(\lambda)\frac{\lambda_{0}}{c}
	\end{equation}
	since the Doppler relation is given by $v/c=(\lambda-\lambda_{0})/\lambda_{0}$, where $c$ is the speed of light. This produces a distribution of the one-dimensional line-of-sight velocity $v$ with the form

	\begin{equation}\label{inter_eq2}
	I(\lambda)\propto \left(1+\frac{(\lambda-\lambda_{0})^{2}}{\kappa 2\sigma_{\kappa}^{2}}\right)^{-\kappa}\rightarrow
	f(v)\propto \left(1+\frac{v^{2}}{\kappa v_{th}^{2}}\right)^{-\kappa}
	\end{equation}
	where we define $v_{th}^{2}=2\sigma_{\kappa}^{2} c^{2}/\lambda_{0}^{2}=2A[3]^{2} c^{2}/\lambda_{0}^{2}$ and where we can interpret $v_{th}$ as the most probable speed of the distribution. At high $\kappa\to\infty$, the kappa distribution tends to a one-dimensional Maxwellian distribution

	\begin{equation}
	f(v)\propto\exp{\left(-\frac{v^2}{v_{th}^{2}}\right)}
	\end{equation}
	 where $v_{th}=\sqrt{2k_{B}T/M}$ is the thermal speed of the distribution.  At small $\kappa$ and high $v$, the kappa distribution resembles the form of a power law $f(v)\sim v^{-2\kappa}$ \citep[cf][]{2014ApJ...796..142B,2009JGRA..11411105L}.

	There are a number of different forms and interpretations of the kappa distribution in the literature, e.g. first and second kinds, and forms where either the ``temperature" or ``thermal velocity" is kappa dependent e.g. \citet[][]{2016arXiv160204132L,2015JGRA..120.1607L,2014Entrp..16.4290L,2009JGRA..11411105L,2009PhPl...16i4701H,2004PhPl...11.1308L}. The physical interpretation of the value of $\kappa$ obtained from the line fitting, in terms of the ion velocity distribution, therefore depends on the form of the ion kappa distribution used. We are only fitting for the distribution of line-of-sight velocity, and this leads to the form in Equation \ref{inter_eq2} with index $\kappa$ (and the corresponding 3-D velocity distribution will have index $-\kappa - 1$) but other forms of kappa distribution for the 3-D velocity have different indices, e.g. \citet{2014ApJ...796..142B}. Nonetheless we can say that the small $\kappa$ values that we find indicate a non-thermal ion distribution regardless of the precise form of the kappa distribution used. We can also comment on the flare thermal environment.

	In the plasma, the isotropic ion motions are responsible for the line shape, which will depend on their relative speed, but the electrons are responsible for the existence of Fe XVI (or any line emission). Hence for the formation of Fe XVI, we only care about the interaction of the Fe ions with the electrons. We know that Fe XVI is formed mainly at electron temperatures of $2$ to $4$ MK, so at $\sim3$ MK, the ion speed (at $\sim3$ MK) is $v_{th}\sim30$ km/s, while the electron thermal speed at this temperature is of the order $10^{3}-10^{4}$ km/s. Therefore, the presence of an accelerated non-thermal heavy ion distribution, even with velocites of 10$\times$ to 100$\times$ the thermal speed, should have very little effect on the formation of Fe XVI, and its emission during the flare.

	\begin{figure*}
	\centering
	\includegraphics[width=17cm,angle=0]{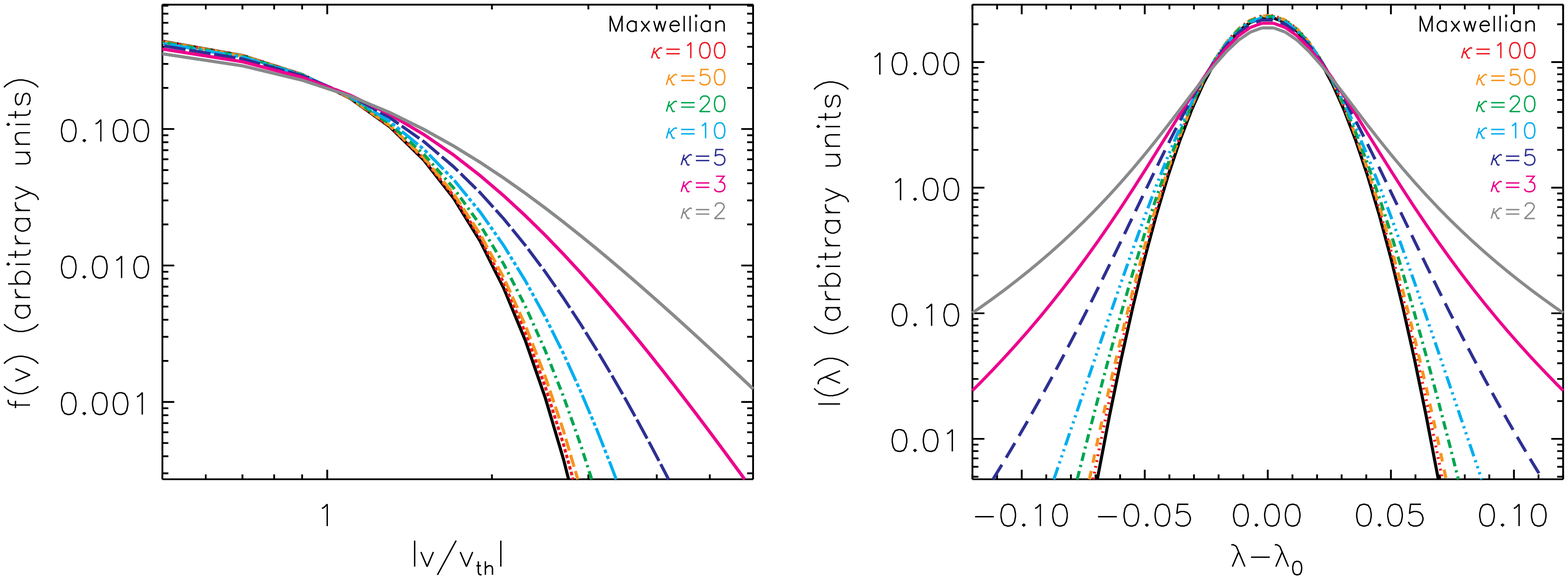}
	\includegraphics[width=17cm,angle=0]{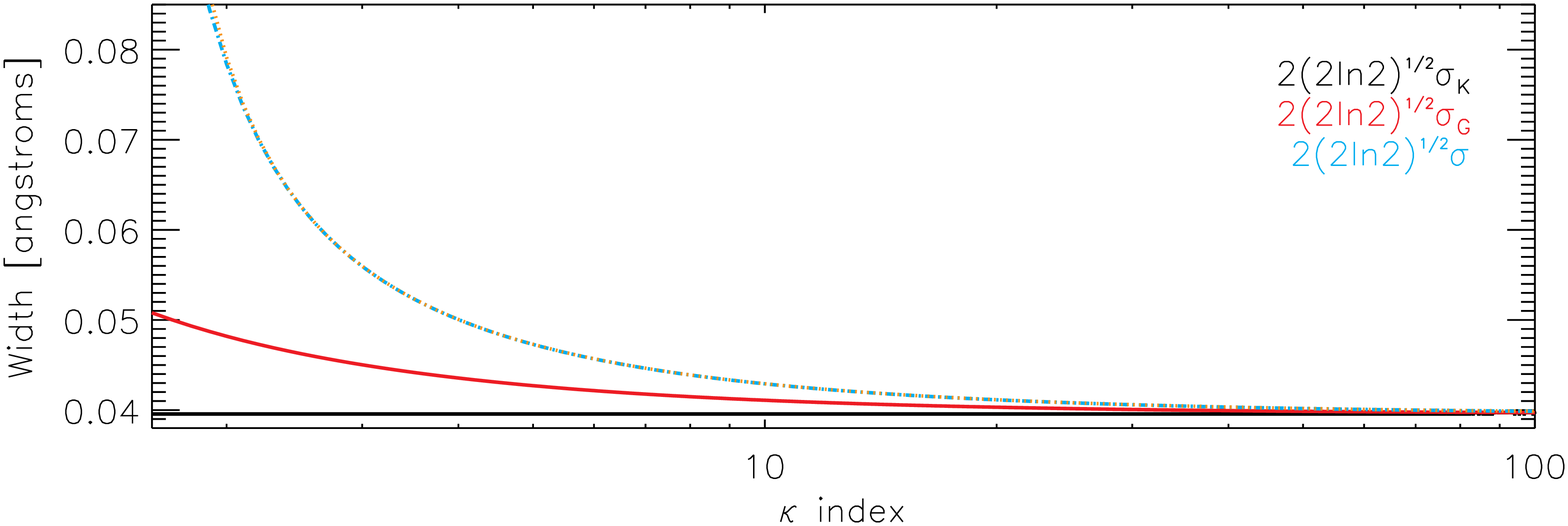}
	\caption{Top left: The line of sight ion velocity distribution $f(v)$ plotted against the absolute value of velocity $|v/v_{th}|$, for a Maxwellian distribution and seven different kappa distributions using $\kappa=100,50,20,10,5,3,2$, using Equation \ref{inter_eq2}. Top right: The corresponding spectral lines produced from each of the ion velocity distributions in the left panel. We use the Fe XVI parameters of $\log_{10}{T}=6.4\sim2.5$ MK, iron mass $M=56\times1.673\times10^{-24}$ g, $\lambda_{0}=262.9760~$\AA and set $n=1$ cm $^{-3}$. The line profile curves all have the same value of $\sigma_{\kappa}$ and $v_{th}$, but different values of $\kappa$. Bottom: $2\sqrt{2\ln{2}}\sigma_{\kappa}$ (black) plotted against $\kappa$ index. The width is displayed as a `Gaussian FWHM' for easy comparison with a Gaussian line profile. Each kappa line profile is fitted with a single Gaussian and the Gaussian widths are also plotted ($2\sqrt{2\ln{2}}\sigma_{G}$ - red curve) for comparison. The actual line distribution variance $\sigma^{2}$ is also shown (displayed as $\sigma \times 2\sqrt{2\ln{2}}$) and calculated using $\sigma^{2}=\sigma_{\kappa}/\left(1-3/2\kappa \right)$ (light blue).}
	\label{kap_fig}
	\end{figure*}

	The timescales $\tau$ for energy exchange vary between species \cite[e.g.][]{1981phki.book.....L}:
	\begin{equation}\label{tau}
	\tau_{ee}^{E}:\tau_{ii}^{E}:\tau_{ei}^{E}\sim1:\sqrt{\frac{m_{i}}{m_{e}}}:\frac{m_{i}}{m_{e}}.	
	\end{equation}
	Since the Fe-to-electron mass ratio is $\sim1836.15\times56$, electrons are the quickest to form thermal equilibrium in a warm target via Coulomb collisions \citep[cf][]{2015ApJ...809...35K,2015JPhCS.642a2013J} with each other. Heavy Fe ions would take $\sim319\times\tau_{ee}^{E}$ to equilibrate with each other but over $10^{5}\times\tau_{ee}^{E}$ to form thermal equilibrium with the electrons. The form of Equation \ref{inter_eq2} can provide an ion thermal velocity $v_{th}=\sqrt{2k_{B}T/M}$ with temperature $T$ that the ions would have in equilibrium with a background population, as $\kappa\to\infty$, without acceleration i.e. the temperature of an initial Maxwellian distribution from which ions were accelerated, not the mean energy of the accelerated ion distribution. We can visualise a simple scenario where an accelerated heavy ion distribution is embedded within a sea of thermalised electrons at temperature $T$. From the observations, we can estimate such temperatures. Calculating $v_{th}$ simply from the kappa-Gaussian fit parameter $\sigma_{\kappa}=A[3]$ (and setting $\kappa\rightarrow\infty$) gives $T=M(c\sigma_{\kappa}/\lambda_{0})^2/k_{B}$, and we find in regions 1, 3, 4 and 7, $T=$ 4.1 MK, 4.5 MK, 4.2 MK and 3.6 MK. Fe XVI could easily form at these temperatures (see Figure \ref{fig_goft}). 

	We also estimate an ion temperature from the single Gaussian fit, by removing the instrumental broadening only. From the Gaussian fitting, the observed line widths are $\sim0.06~$\AA to $0.07~$\AA. If the width is assumed to be completely due to isothermal motions and converted to a temperature $T$, then we calculate temperatures that are approximately equal to or higher than $6$ MK. At these temperatures, the contribution function is an order of magnitude below that of the peak $G(T)$ values (see Figure \ref{fig_goft}), and Fe XVI is less likely to form. This is why the concept of an `excess' broadening is often used in line broadening studies. If the line profile is a kappa line shape with a low $\kappa$ index, then a Gaussian fit tries to account for the broad wings by fitting a Gaussian with a larger $\sigma_{G}$, leading to an excess width. An example of this is shown Figure \ref{kap_fig}, bottom panel.

	From this discussion we speculate that it is possible that the total excess broadening in this case (found from Gaussian fitting) and the overall non-Gaussian Fe XVI line profiles can be formed entirely by the presence of an accelerated non-thermal Fe ion population with a background electron temperature of 3.5 to 4.5 MK. However, more work is required to discuss this speculation further. Also, in such a scenario the presence of other macroscopic plasma motions is not required. However, their existence cannot be ruled out completely by the study. Therefore, the scenario envisaged in this section and other processes is the subject of ongoing work.

	\section{Conclusions}\label{conclusions}
	In this paper we have used {\it Hinode} EIS to investigate solar flare Fe XVI ($262.976~$\AA) spectral line profiles at a single time interval covering the HXR and SXR peaks, in different regions of an X-class flare. We studied seven flare regions, from the HXR-emitting footpoints to the coronal loop-top source, and we were able to investigate non-Gaussian profiles in all but one region. We showed that the presence of non-Gaussian line shapes can be detected using {\it Hinode} EIS and suitable lines with a low level of noise ($\sim <10 \%$) and without the present of obvious directed mass motions (a low level of skewness). Two independent investigations suggested that the Fe XVI lines emitted during the flare were better described by non-Gaussian line shapes. Taking into account the uncertainties, a higher-moments analysis found kurtosis values greater than 3, even with the presence of a broad Gaussian EIS instrumental profile, and a small studied wavelength range of $\lambda_{0}\pm0.2~$\AA. This suggested that some of the lines were more peaked with broader wings than a Gaussian profile. Suitable lines were fitted with single Gaussian line profiles and single convolved kappa-Gaussian line profiles. The convolved kappa-Gaussian profile accounts for a broad instrumental profile (Gaussian part) and a physical profile (kappa part). Compared to the single Gaussian fits, the kappa-Gaussian convolved profiles (and single kappa profiles that approximate the overall line shape) produced the lowest reduced $\chi^{2}$ values. The kappa-Gaussian convolved profiles were able to fit the peaked lines with broad wings. We found conclusive evidence of non-Gaussian profiles in the loop-top (region 1), northern HXR footpoint (3 and 4) and southern ribbon (7) regions. In loop leg regions (2 and 5), the Gaussian fit reduced $\chi^{2}$ values were also very low and close to 1. Hence, the line profile shapes in the loop leg regions could not be confidently determined. Our investigation is mainly in agreement with \citet[published but not yet refereed][]{2013arXiv1305.2939L}, that found evidence of non-Gaussian line profiles in non-flaring regions.
	
	We briefly discussed one possible interpretation of non-Gaussian Fe XVI line profiles, accelerated non-thermal, isotropic ion populations, that can be described with a kappa velocity distribution. The kappa values found from the convolved kappa-Gaussian fits were very low, between 2 and 3.3. Such values correspond to highly accelerated ion distributions far from a Maxwellian ion population (see Figure \ref{kap_fig}). The high velocity power-law part of the kappa distribution given by $f(v)\sim v^{-2\kappa}$, gives spectral indices of 4 to 6.6 for a one dimensional velocity distribution. We suggested that the line shape and total broadening (including the excess broadening) of Fe XVI ($262.976~$\AA), in this flare, could be completely explained by the presence of a non-thermal ion population within a background thermalised electron population with $T\sim4$ MK. 

	The properties of accelerated solar flare non-thermal electrons are routinely deduced by bremsstrahlung X-ray observations \citep[see recent reviews by][]{2011SSRv..159..301K,2011SSRv..159..107H}, currently using X-ray imaging and spectroscopy provided by the Ramaty High Energy Solar Spectroscopic Imager (RHESSI) \citep{2002SoPh..210....3L}, but the mechanism(s) responsible for their acceleration, and in many other astrophysical scenarios, still remains poorly understood \citep[e.g.][]{2011SSRv..159..357Z}. The properties of solar flare-accelerated protons and heavier ions can be studied using keV to MeV gamma-ray bremsstrahlung continuum and line emission \citep[see e.g.][]{2011SSRv..159..167V}. However, it is rare for such high photon energies to be detectable during the majority of flares, and hence the form of the accelerated protons and heavier ions are usually unknown during the flare. Investigating the presence of non-Maxwellian ion populations from suitable EUV spectral lines, may provide a possible method of studying lower energy accelerated ion populations from abundant EUV and UV observations. Such flare observations could help to distinguish between different competing acceleration mechanisms or constrain the parameters of a particular acceleration mechanism. Using the non-thermal ion interpretation, the next step is to compare the heavy ion spectra with the electron distribution spectra from X-ray observations. Although not discussed, other processes such as random bulk plasma motions could also be responsible for the non-Gaussian line profiles and broadening, and the analysis of the line shape could provide the velocity distribution of bulk plasma motions. Examining other flares at different heliocentric angles on the solar disk might help to distinguish between the different possible processes.

	We also note that we see non-Gaussian line profiles at all flare times from $\sim$01:25 UT to 02:30 UT, but a full temporal study was beyond the scope of this paper. This is the subject of ongoing work. Overall, we have shown that it is possible to use EIS data for line profile studies. We hope to continue the study with other EIS spectral lines and with higher spectral resolution it IRIS data. Fe XVI is also present before and after the flare, but at much lower intensities (lower by an order of magnitude), making a line profile analysis more challenging. Therefore, it is difficult to investigate if the non-Gaussian line shapes are actually produced or changed by the onset of the flare. Finally, in this study, we cannot rule out that the non-Gaussian line profiles are a product of the instrumental profile but this is the topic of continuing work.

	\begin{acknowledgements}
	NLSJ, LF and NL gratefully acknowledge the financial support by STFC Consolidated Grant ST/L000741/1.
	CHIANTI is a collaborative project involving George Mason University, the University of Michigan (USA) and the University of Cambridge (UK). Hinode is a Japanese mission developed and launched by ISAS/JAXA, collaborating with NAOJ as a domestic partner, NASA and UKSA as international partners. Scientific operation of the Hinode mission is conducted by the Hinode science team organized at ISAS/JAXA. This team mainly consists of scientists from institutes in the partner countries. Support for the post-launch operation is provided by JAXA and NAOJ (Japan), UKSA (U.K.), NASA, ESA, and NSC (Norway). The research leading to these results has received funding from the European Community's Seventh Framework Programme (FP7/2007-2013) under grant agreement no. 606862 (F-CHROMA). We thank the referee for helping to improve the paper, particularly regarding the possible instrument effects, and Eduard Kontar for useful comments.
	\end{acknowledgements}

	\bibliographystyle{aa}
	\bibliography{arvix_new_nj}

	\end{document}